\documentclass{article}
\usepackage[utf8]{inputenc}
\usepackage{amsmath}
\usepackage{amssymb}
\usepackage[normalem]{ulem}
\usepackage{cancel}
\usepackage[dvipsnames]{xcolor}
\usepackage{stackengine}
\usepackage{graphicx,caption}
\usepackage{multirow}
\usepackage{pgfplotstable}
\usepackage{physics}
\usepackage{soul}
\usepackage[font=small]{caption}

\definecolor{myurlcolor}{rgb}{0,0,0.7}
\definecolor{myrefcolor}{rgb}{0.8,0,0}
\usepackage[unicode=true,pdfusetitle, bookmarks=false,bookmarksnumbered=false,
bookmarksopen=false, breaklinks=false,pdfborder={0 0 0},backref=false,
colorlinks=true, linkcolor=myrefcolor,citecolor=myurlcolor,urlcolor=myurlcolor]
{hyperref}

\usepackage{hyperref}
\DeclareMathOperator*{\argmax}{arg\,max}

\usepackage[
backend=bibtex,
sorting=none
]{biblatex} 
\usepackage{url}

\newtheorem{definition}{Definition} 

\title{Automated Gadget Discovery in Science}

\author{\small Lea M. Trenkwalder$^1$\footnote{These authors contributed equally to this work.} , Andrea López Incera$^1$\protect\CoAuthorMark \, , Hendrik Poulsen Nautrup$^1$,\\ \small Fulvio Flamini$^1$, and Hans J. Briegel$^{1,2}$ }
\date{%
    \small$^1$Institute for Theoretical Physics, University of Innsbruck, 6020 Innsbruck, Austria\\%
    $^2$Department of Philosophy, University of Konstanz,  78457 Konstanz, Germany\\[2ex]%
    \large
    \today
}
\newcommand\CoAuthorMark{\footnotemark[\arabic{footnote}]} 

\begin{document}
\maketitle

\abstract{In recent years, reinforcement learning (RL) has become increasingly successful in its application to science and the process of scientific discovery in general. However, while RL algorithms learn to solve increasingly complex problems, interpreting the solutions they provide becomes ever more challenging. In this work, we gain insights into an RL agent’s learned behavior through a post-hoc analysis based on sequence mining and clustering. Specifically, frequent and compact subroutines, used by the agent to solve a given task, are distilled as gadgets and then grouped by various metrics. This process of gadget discovery develops in three stages: First, we use an RL agent to generate data, then, we employ a mining algorithm to extract gadgets and finally, the obtained gadgets are grouped by a density-based clustering algorithm. We demonstrate our method by applying it to two quantum-inspired RL environments. First, we consider simulated quantum optics experiments for the design of high-dimensional multipartite entangled states where the algorithm finds gadgets that correspond to modern interferometer setups. Second, we consider a circuit-based quantum computing environment where the algorithm discovers various gadgets for quantum information processing, such as quantum teleportation. This approach for analyzing the policy of a learned agent is agent and environment agnostic and can yield interesting insights into any agent’s policy.}


\section{Introduction}\label{sec:introduction}

From playing games to predicting the structure of proteins~\cite{Vinyals2019, Jumper2021, Schrittwieser2020}, deep reinforcement learning (DRL) has become immensely successful in solving complex problems. This rapid progress has inspired the application of DRL in various scientific disciplines, ranging from biology \cite{Zhou2021} and medicine \cite{Mahmud2017} to physics \cite{Fosel2021, Ostaszewski2021}. These applications focus specifically on acquiring novel or optimized solutions to specific problems. In contrast, machine-assisted scientific
discovery goes beyond finding an optimal solution to a narrowly defined task and asks for generic methods that can assist scientists in basic research~\cite{Nautrup2022,Iten2020, Melnikov2018, Ried2019, Wu2019, DeSimone2019,DAgnolo2019, Rahaman2019,Thiede2020, Krenn2021}. As such, problem solving is only one part of the scientific discovery process and can be
complemented by methods such as explainable reinforcement learning (XRL) to improve the transparency of black box algorithms. In this way, we provide human users with tools to
interpret complex RL policies and even learn from them.

ML-assisted research requires systems that deliver insights into the solution strategy. For example, elements extracted from the observation of the agent's behavior can inform non-expert users about the strengths and limitations of the reinforcement learning (RL) agent \cite{Sequeira2020}. Often enough the path towards a result is as interesting as the result itself. This is especially true for problems in quantum physics, where our intuition easily fails to grasp the strategy that led to the desired result. In Ref.~\cite{Melnikov2018}, for example, an active learning agent learns to prepare highly entangled quantum states in simulated quantum experiments. Interestingly, in an expert post-hoc analysis, the authors (re-)discovered multiple tools (so-called gadgets) that were developed by the agent for entanglement generation that have since been used in actual quantum optics experiments \cite{Malik2016}. Similarly, in the groundbreaking work where a DRL agent beats the world champion in Go, a post-hoc expert analysis reveals the use of previously unknown game moves \cite{Silver2017}. We see such tools and moves as instances of \textit{gadgets}: frequent and compact sequences of actions that are useful to solve certain RL tasks.

\begin{figure}[htbp]
\begin{center}
\includegraphics[width=14cm]{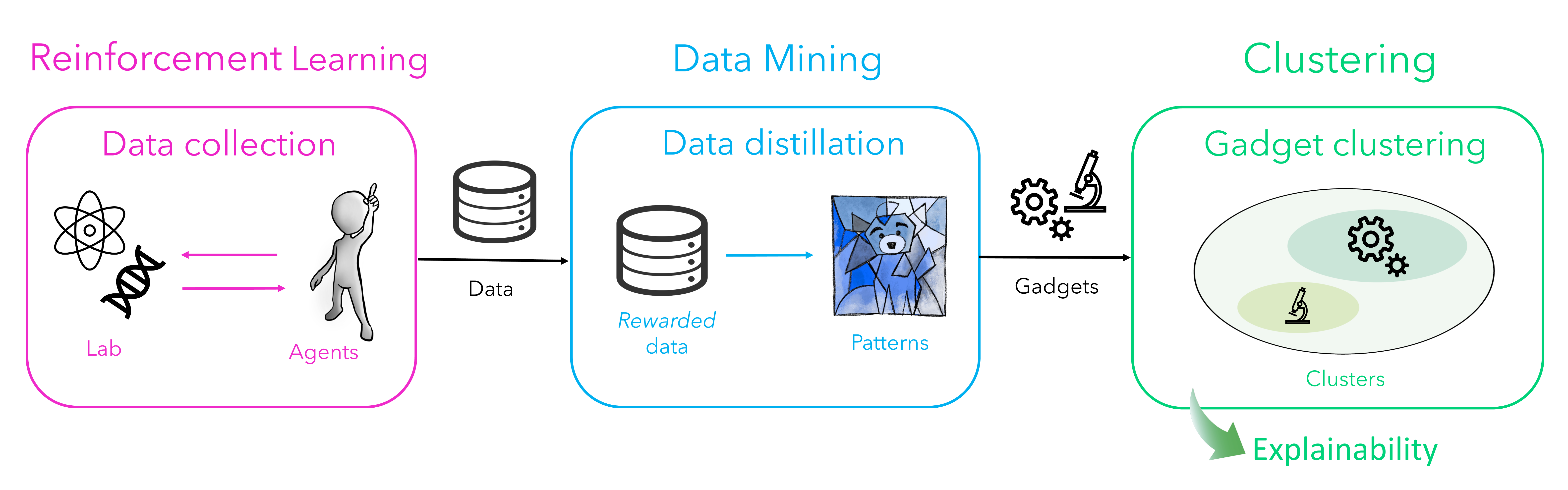}
\caption{\textbf{Architecture for automated gadget discovery.} The algorithm consists of three main parts: reinforcement learning (RL), data mining (DM), and clustering. Data generated by the RL algorithm (see Sec.~\ref{sec:methods:RL}) is analyzed by a DM algorithm that distills gadgets (see Sec.~\ref{sec:methods:data_mining}). A clustering algorithm groups the gadgets by environment-specific information, by utility, or by the specific context that triggered the use of the gadget (see Sec.~\ref{sec:methods:clustering}). Selected gadgets are then presented in a human-readable form that improves interpretability as defined in Sec.~\ref{subsec:explain}}
\label{fig:architecture_outline}
\end{center}
\end{figure}

In this paper, we introduce a framework for automated gadget discovery in RL based on intrinsic motivation, data mining techniques, and clustering (see Fig.~\ref{fig:architecture_outline}). This framework is a post-hoc XRL \cite{Puiutta2020} approach in discrete state and action spaces, where gadgets provide human-readable information about the agent's behavior.
Gadgets distillation occurs in three stages (see Fig.~\ref{fig:architecture_outline}): First, we use RL agents to explore a task environment. Then, we employ a data mining subroutine~\cite{Zhou2016,Zaki2001} to extract gadgets from the action sequences.
Lastly, to simplify the post-hoc analysis, we use a density-based clustering method~\cite{Campello2013} to group gadgets by useful criteria. The output gadgets contain relevant information about the task environment. As such the gadgets serve as instructive examples for the user to better understand the task environment and policy.

We demonstrate our method on two tasks inspired by problems in quantum physics. The first RL task is the quantum optics environment studied in Ref.~\cite{Melnikov2018}, where we show that our architecture discovers all the same gadgets. The second RL task is a quantum circuit environment where the goal is to preserve quantum information. Here, we show the capabilities of our method through a comparative post-hoc expert analysis of the extracted gadgets. For example, the proposed architecture discovers gadgets that teleport the quantum information and protect it from measurements, and it groups them separately from other simpler gadgets.

The paper is organized as follows. In Sec.~\ref{sec:methods}, we introduce the architecture and the main methods used in each stage of the algorithm (see Fig.~\ref{fig:architecture_outline}). We then test our architecture on two RL environments in Sec.~\ref{sec:results}.

\section{Methods}\label{sec:methods}
In this section, we will introduce the main methods used in the proposed architecture (see Fig.~\ref{fig:architecture_outline}). 
With this approach, we develop a XRL method for science by (i) collecting data from the policy of an RL agent (see Sec.~\ref{sec:methods:RL}), (ii) extracting gadgets from the RL data (see Sec.~\ref{sec:methods:data_mining}), and (iii) cluster the gadgets (see Sec.~\ref{sec:methods:clustering}). In this way, we improve the interpretability of both the gadgets and the environment for the human user. 
\subsection{Reinforcement learning}\label{sec:methods:RL}
In reinforcement learning (RL) an agent interacts with its environment. 
At each time step $t$, the agent perceives information about the state of the environment in the form of an observation $o\in O$, takes an action $a \in A$, and receives feedback in the form of a reward $r\in \mathbb{R}$. In this work, we consider the action set $A$ and the observation set $O$ to be finite and discrete sets. An episode describes all interactions between an agent and its environment until a termination condition is fulfilled. To simplify the interpretability of interaction sequences, we assume feature encodings for both actions and observations. That is, an action $a$ can be described by $n$ features $a=(f_1,f_2,...,f_n)$. For example, if actions describe movements, their features may be direction and speed. Even with minimal knowledge of the environment, such an encoding is typically easy to define. At the same time, this encoding gives rise to two complementary ways of clustering gadgets (see Sec.~\ref{sec:methods:clustering}) and allows for more control of the mining procedure (see Sec-\ref{subsubsec:postprocessing}).

In RL, the agent's behavior is described by a conditional probability distribution called \textit{policy} $\pi(a|o)$. The agent's interaction with the environment is captured by a sequence of observations, actions and rewards. A standard figure of merit to evaluate the agent's performance is given by the \textit{return}, which is the sum of rewards discounted by an environment-specific discount factor $\gamma$. For our purpose, different from standard formulations of the return (see Appendix \ref{appendix:DDQN}), we define it as:
\begin{align}\label{eq:reward_rl} 
G(d)=\sum_{j=0}^{T-1} r_{j}\gamma^{j+1}, 
\end{align}
where $d=(o_0,a_0,r_0,...,o_{T-1},a_{T-1},r_{T-1})\in \left( O\times A\times \mathbb{R} \right)^T$ is the interaction sequence and $T$ is a fixed length called \textit{horizon}. The goal of the agent is to find the optimal policy which maximizes the expected return.\\
Scientific tasks can be phrased as RL problems with their solutions encoded in the agent's policy. This technique has already been applied successfully in various fields. In this work, we analyze the obtained policy by extracting gadgets, i.e.: frequent and compact subsequences of actions with large returns (see 
Sec.~\ref{sec:methods:data_mining}). To extract gadgets, we collect sequences $d$ of fixed length $T$ with return values $G(d)$ larger than a threshold $G_\textrm{min}$. These gadgets provide a way to interpret and explain the agent's policy, ultimately allowing us to draw conclusions about the structure of the environment. This is particularly useful in environments such as the one discussed in Ref.~\cite{Melnikov2018}, where the authors can conclude that certain interferometers are particularly useful for generating quantum entanglement.\\ To ensure that we gain as much information about the environment as possible, we employ a exploration-driven RL algorithm as opposed to the usual exploitation-exploration-balanced approach. That is, while large returns are important, we prioritize a diverse action sequence set. For example, we use an agent with intrinsic motivation (IM) that receives internal rewards to facilitate exploration. We refer to Appendix~\ref{appendix:rl} for a consideration of the different types of RL environments and how they facilitate the use of different RL agents. Independent of the choice of RL agent, gadgets are extracted from the collected dataset by the data mining algorithm described in the following subsection. 


\subsection{Gadget mining}\label{sec:methods:data_mining}

In this work, we consider gadget mining as an extension of sequence mining to interaction sequences. Pattern mining, in turn, consists in discovering interesting and possibly useful patterns in a database. When pattern mining has to take into account the temporal order of events, it is called sequential pattern mining, or sequence mining for short~\cite{Fournier-Viger2017}. In this work, we rank subsequences of labeled, sequential data as patterns by their \textit{support}, \textit{cohesion}, and \textit{interestingness}, following the approach and notation introduced in Ref.~\cite{Zhou2016}. 

\subsubsection{Support, cohesion, and interestingness}\label{subsec:int}
Consider a dataset $D$ consisting of $|D|$ sequences. Each sequence in the dataset $D$ can be described as a sequence of events $e = (i,\tau)$, where $i\in J$ is an item and $\tau \in \mathbb{N}$ is the corresponding time stamp. $J$ denotes the set of all items that can arranges as a sequence. A sequence of $l$ events is denoted as $d = (e_1, e_2,...,e_l)$. A sequence $\tilde{d}= (\tilde{e_{1}}, \tilde{e_{2}},...,\tilde{e_{m}})$ is a subsequence of the sequence $d$, denoted as $\tilde{d}\sqsubseteq d$, if there exist $m$ indices $j_k$ for which $1\leq j_k<j_{k+1}\leq l$ such that $\tilde{e_{k}}=e_{j_k}$ for all $k \in \{1,...,m\}$. Subsequences that start at timestamp $\tau_s$ and end at timestamp $\tau_e$ are denoted as $\tilde{d}_{(\tau_s,\tau_e)}$.

Following Ref.~\cite{Zhou2016}, the support $F_D(\tilde{d})$ of a subsequence $\tilde{d}$ quantifies the frequency of sequences $d$ that contain the subsequence $\tilde{d}$ within the dataset $D$:
\begin{align}
    F_D(\tilde{d}):= \frac{|N_D(\tilde{d})|}{|D|},
\end{align}
where set $N_D(\tilde{d})= \{d\in D|\tilde{d} \sqsubseteq d\}$ is the set of all sequences $d$ that contain $\tilde{d}$. The cohesion quantifies the average compactness of a subsequence within a dataset $D$ by:
\begin{align}
   C_D(\tilde{d}):= \frac{|\tilde{d}|}{\overline{W}_D(\tilde{d})},
\end{align}
where the average shortest interval length $\overline{W}_D(\tilde{d})$ is determined by $\overline{W}_D(\tilde{d})=\frac{\sum_{d\in N_D(\tilde{d})} W(d,\tilde{d})}{|N_D(\tilde{d})|}$, where the shortest interval length $W(d,\tilde{d})$ of a subsequence $\tilde{d}$ in $d\in N_D(\tilde{d})$ is defined as $W(\tilde{d},d)= \min\{\tau_e-\tau_s+1|\tau_s\leq \tau_e \land \tilde{d} \sqsubseteq d_{(\tau_s,\tau_e)}\}$. 

With these definitions, the interestingness $I(\tilde{d})$ of a subsequence $\tilde{d}$ is then simply the product of its support and its cohesion:
\begin{align}
   I_D(\tilde{d}):=F_D(\tilde{d})\cdot C_D(\tilde{d}).
\end{align}
In the following, we always consider that the support $F_D$, cohesion $C_D$, and interestingness $I_D$ are defined with respect to the full input dataset $D$, which allows us to ease the notation introduced in~\cite{Zhou2016} by omitting the subindex $D$.

\subsubsection{Patterns and gadgets}\label{subsubsection:Patterns and gadgets}

Patterns of a sequential dataset $D$ are defined as subsequences that have a sufficiently large minimal support, cohesion, and interestingness with respect to the dataset:

\begin{definition}[Pattern]
Consider a dataset $D$ of size $|D|$ of sequences $d$ of fixed length $T$ where no sequence has a duplicate. The  sequence $\tilde{d}$ is a subsequence for some $d\in D$. Given a minimum support threshold $F_\textrm{min}\in[0,1]$, a minimum cohesion threshold $C_\textrm{min}\in[0,1]$ and a minimum interestingness threshold $I_\textrm{min}\in[0,1]$, the subsequence $\tilde{d}$ is considered a \textrm{pattern} w.r.t. $D,F_\textrm{min},C_\textrm{min},I_\textrm{min}$ if $F(\tilde{d})\geq F_\textrm{min}$, $C(\tilde{d})\geq C_\textrm{min}$, and $I(\tilde{d})\geq I_\textrm{min}$. 
\end{definition}

Gadgets are sequences of actions that are also patterns. To define gadgets, we first define the action sequence $d_A\sqsubseteq d$ of an interaction sequence $d$ as its restriction to the action set $A$, i.e., $d_A:=d_{|A}\in A^{T}$.

\begin{definition}[Gadget]\label{def:gadget} Given a return threshold $G_\textrm{min}\in\mathbb{R}$, consider a dataset $D_\textrm{R}$ of rewarded interaction sequences such that $G(d)\geq G_\textrm{min},\: \forall d\in D_\textrm{R}$. Let $D_\textrm{A}:=\{d_{|A}|d\in D_\textrm{R}\}$ be the associated dataset of action sequences. 
A pattern w.r.t. $D_A, F_\textrm{min}, C_\textrm{min}, I_\textrm{min}$ is a \textrm{gadget} w.r.t. $D_A,F_\textrm{min}, C_\textrm{min}, I_\textrm{min}, G_\textrm{min}$.
\end{definition}

Notably, we ignore observations and rewards in the definition of gadgets. However, we will use this information for clustering in the last stage of the algorithm, as described in Sec.~\ref{sec:methods:clustering}. In Appendix~\ref{appendix:dm}, we describe an efficient approach for identifying gadgets based on the SPADE algorithm~\cite{Zaki2001}, which was originally designed to distill patterns from sequential datasets. The algorithm takes as input a dataset $D_\textrm{A}$ of rewarded action sequences and hyperparameters $F_\textrm{min},C_\textrm{min},I_\textrm{min}$ and outputs a set of gadgets, with no predefined elements or length.

\subsubsection{Pre- and postprocessing}\label{subsubsec:postprocessing}
To facilitate the interpretability of gadgets, we use additional data processing steps whenever possible. For example, as mentioned in Sec. \ref{sec:methods:RL}, knowledge about the structure of the action space can be used to encode each action $a$ by a feature vector $a=(f_1,f_2,...,f_n)$. If features possess a hierarchy, this allows controlling the granularity of the mining procedure by removing irrelevant features from the encoding before mining. 
In the same way, we can apply additional filters in postprocessing to remove gadgets with certain undesirable properties or to restrict the total amount of found gadgets.
\subsection{Clustering gadgets}\label{sec:methods:clustering}

The mining algorithm described in Appendix~\ref{appendix:dm} returns gadgets as defined in Sec.~\ref{sec:methods:data_mining}. All gadgets can be associated and ranked according to the three quantities presented in Sec.~\ref{subsec:int}. However, these quantities alone do not provide any relational information about gadgets that could be particularly interesting to a user. For instance, the agent should be able to autonomously group similar gadgets by different criteria. 
We use the well-established, density-based clustering algorithm HDBSCAN~\cite{Campello2013} to cluster gadgets in a metric space induced by two different distance measures. More specifically, clusters are associated with dense regions of gadgets, while gadgets that are farther apart according to these distances are more likely to be associated with different clusters. The algorithm is described in more detail in Appendix~\ref{appendix:HDBSCAN}.
In addition, we also make use of any extra information provided by the environment to refine the clustering process and group gadgets in a more meaningful way.

In the following, we consider two complementary approaches to measuring distances for clustering: The utility-based method described in Sec.~\ref{subsec:utility}, which directly compares gadgets, and the context-based method in Sec.~\ref{subsec:context}, which compares the states of the environment when gadgets are used (context).
Both methods serve to increase the interpretability of the gadgets, as we discuss in Sec.~\ref{subsec:explain} in the broader context of interpretability. 

\subsubsection{Utility-based clustering}\label{subsec:utility}

Utility-based clustering groups gadgets by their utility, which separates the gadgets by their action composition. As defined in Sec.~\ref{subsubsection:Patterns and gadgets}, gadgets consist of sequences of actions that are described by a set of features (see Sec.~\ref{subsubsec:postprocessing}). These features can be manipulated and used to encode the gadgets and define a suitable distance measure, which will be used during the clustering stage. For instance, a simple and convenient encoding could be a tally vector that counts the number of actions of a certain type each gadget $g$ contains. In general, an encoded gadget can be described as a vector $\phi(g) \in \mathbb{R}^{m}$. With this encoding, we can also define a utility weight $\alpha_i \in \mathbb{R}$ for each vector component $i \in \{1,...,m\}$, which weights the importance of a feature with respect to the other components. The encoding of a gadget $g$ weighted by a list of utilities is denoted as $\phi_w(g) = (\,\alpha_i \phi_i(g),...,\alpha_m \phi_m(g) \,)$. A distance measure between two gadgets $M(g_1,g_2)$ is then given by the distance between their weighted encoding vectors $\phi_w(g_1)$ and $\phi_w(g_2)$. Importantly, we remark that the specific encoding, utility weights, and distance measure are all user-defined quantities, which provides an extra degree of freedom that can be especially useful when prior knowledge about the task environment is available. In the following sections, we will return to these definitions and discuss the specific choices made for each environment under investigation.
The resulting clusters, can then be further analysed by determining the most frequent feature composition of the gadgets within the cluster and compare them to the most frequent feature composition of the gadgets in other clusters.

\subsubsection{Context-based clustering}\label{subsec:context}

In this section, we consider clustering gadgets by the context they are used in, that means, by the observations that triggered their use. For each gadget $g=(a^{(1)},a^{(2)},...)$, consider the set of all interaction sequences $D_g$ that contain $g$ as a subsequence, i.e. $D_g:=\{d\in D| g\sqsubseteq d\}$. 
Recall from Sec.~\ref{sec:methods:RL} that an interaction sequence is given by $d=(o_0,a_0,r_0,...,o_{T-1},a_{T-1},r_{T-1})\in \left( O\times A\times \mathbb{R} \right)^T$.
The set of all observations that appear prior to the placement of the gadget in the interaction sequence is called \textit{initialization set} $\mathcal{I}_g=\{o_{t} \in d \, \forall d \in D_g| a_t=a^{(1)} \land a_k \neq a^{(1)} \, \forall k < t \}$. As its name suggests, this set contains the observations (the context) that initialized the use of the gadget. 


To cluster gadgets $g_1,g_2$ by their context, we need to define a distance between two initialization sets. We consider two observations $o,q$, consisting of sequences of items (in this paper items will be actions), and their distance $m(o,q)$, which is defined by the user (we use the Hamming distance, as described in Appendix~\ref{appendix:qi_env}). Then, we define the distance between initialization sets as,
\begin{align}\label{eq:distance_initsets}
    M(\mathcal{I}_{g_1},\mathcal{I}_{g_2})=\frac{1}{|\mathcal{I}_{g_1}|+|\mathcal{I}_{g_2}|}\left(\sum_{o\in\mathcal{I}_{g_1}}\underset{q\in\mathcal{I}_{g_2}}{\min}\left( m(o,q)\right)+\sum_{q\in\mathcal{I}_{g_2}}\underset{o\in\mathcal{I}_{g_1}}{\min}\left( m(o,q)\right)\right).
\end{align}
This distance is the average minimum difference between all elements of the two initialization sets. This definition ensures that $M(\mathcal{I}_{g_1},\mathcal{I}_{g_2})=0$ if $\mathcal{I}_{g_1}=\mathcal{I}_{g_2}$. Here, we assumed that $|\mathcal{I}_{g_1}|=|\mathcal{I}_{g_2}|$, which is enforced by setting a maximum size to all initialization sets. 




\subsubsection{Towards explainability}\label{subsec:explain}

As the autonomy and complexity of learning algorithms increase, the need for explainable approaches, that are both transparent and interpretable, become progressively more important. Similarly, in the context of AI-driven science, we deem transparency and interpretability as vital. If AI systems are used as research assistant systems, they must be able to provide the human user with insight into their solution process. 
Despite, the relevance of the term interpretability for artificial intelligence (AI) research, there is no commonly agreed upon definition for it \cite{Puiutta2020}. 
In this work, we adopt the notion of \textit{interpretability} introduced in Ref.~\cite{Puiutta2020}:``the ability to not only extract or generate explanations for the decisions of the model but also to present this information in a way that is understandable by human (non-expert) users to, ultimately, enable them to predict a model’s behavior.''
The XRL method introduced in this paper extracts gadgets from RL data, clusters, and presents them to the user as instructive examples of the agent's behavior in a given task.
This explainability method classifies as post-hoc~\cite{Puiutta2020}, since it is an additional method that is applied after the learning process has finished.

\section{Results}\label{sec:results}

\begin{figure}[htbp]
\begin{center}
\includegraphics[width=0.95\textwidth]{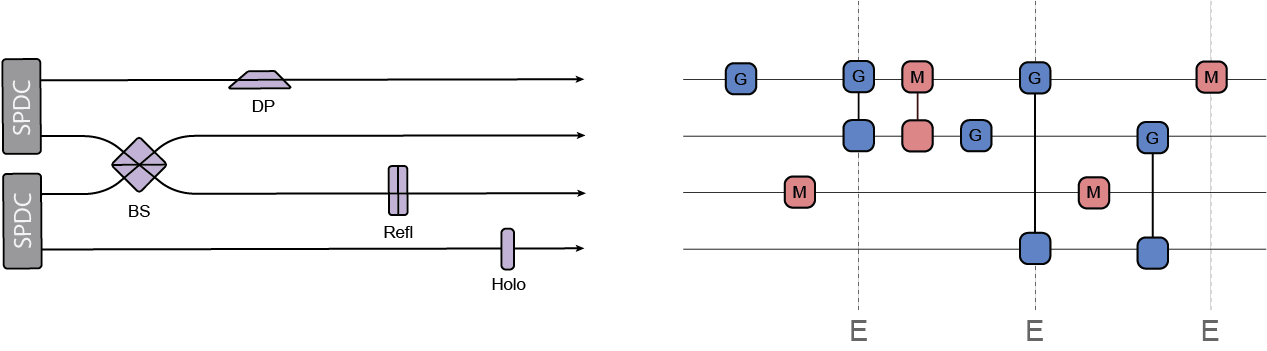}
\caption{\textbf{Quantum optics and quantum information environments.} Left) In the quantum optics (QO) environment, an agent observes an optical setup and places new elements to generate multipartite high-dimensional entanglement in the orbital angular momentum degree of freedom of photons. The initial state consists of two entangled photon pairs produced by spontaneous parametric down-conversion (SPDC). The set of available optical elements consists of beam splitters (BS), Dove prisms (DP), mirrors (Refl), and holograms (Holo). 
Right) In the quantum information (QI) environment, an agent observes a quantum circuit and places circuit elements to protect the information encoded in the first (top) qubit from measurements. The elements placed on the dotted lines marked with the letter E correspond to the initial circuit randomly generated by the environment in this example. The actions of the agent fill up the empty slots (or layers) between the elements of the initial circuit. The set of available actions consists of various single-qubit rotations and two-qubit entangling gates (G), as well as, single-qubit measurements and two-qubit Bell measurements (M).}
\label{fig:envs}
\end{center}
\end{figure}

In this section, we demonstrate our architecture in two quantum-inspired RL environments (see Fig.~\ref{fig:envs}). The reason for this choice is that quantum physics is a field of science where human intuition is often challenged or fails. Therefore, an automated procedure that does not depend on our biased idea of reality may be of help for distilling key properties of the RL environment. In Sec.~\ref{sec:results_qo}, we start from the environment from Ref.~\cite{Melnikov2018} and use an intrinsically motivated (IM) RL agent to design quantum optics experiments that produce multi-partite high-dimensional entanglement. In Sec.~\ref{sec:results_qi}, we consider a quantum information environment where a double deep Q-network~\cite{VanHasselt2015} is tasked to design quantum circuits that preserve quantum information. For each task, we describe the environment and agent in more detail before we present the results of the gadget mining and clustering. A summary of the two environments is given in Table~\ref{table: summary}.

\begin{table}[]
\centering
\renewcommand{\arraystretch}{1.4}
\resizebox{\columnwidth}{!}{%
\begin{tabular}{cccccc}
\hline
\textbf{Environment} & \textbf{Agent} & \textbf{Task}  & \textbf{Start}  & \begin{tabular}[c]{@{}c@{}}\textbf{Clustering} \\ \textbf{method}\end{tabular} & \begin{tabular}[c]{@{}c@{}}\textbf{Clustering} \\ \textbf{based on}\end{tabular}                                                 \\ \hline

\multirow{2}{*}{\textbf{QO}} & \multirow{2}{*}{MCTS+IM} & \multirow{2}{*}{Create entanglement}    & \multirow{2}{*}{Fixed}    & Utility   & Type of operations \\ \cline{5-6} 
                      & &  & & SRV  & Final outcome \\ \hline
                      
\multirow{2}{*}{\textbf{QI}} & \multirow{2}{*}{DDQN}    & \multirow{2}{*}{Preserve qu. information} & \multirow{2}{*}{Random} & Utility         & Type of operations \\ \cline{5-6} 
                      &  & & & Context & Prior observation \\ \hline

\end{tabular}
}
\renewcommand{\arraystretch}{1}

\caption{Summary of the main differences between the two RL environments we consider to test the architecture presented in Sec.~\ref{sec:methods}. The results for the quantum optics (QO) environment, where a Monte Carlo tree search (MCTS) algorithm with intrinsic motivation (IM) is applied, are discussed in Sec.~\ref{sec:results_qo}. In this environment, each quantum state has an associated Schmidt-rank vector (SRV). The results for the quantum information (QI) environment, where a double deep Q-learning (DDQN) agent is used, are presented in Sec.~\ref{sec:results_qi}} \label{table: summary}

\end{table}

\subsection{Discovering quantum optics gadgets}\label{sec:results_qo}

In this section, we apply our algorithm to discover quantum optics gadgets, in experimental setups that challenge our human intuition based on classical physics. 

\subsubsection{Quantum optics environment}

In this RL environment, an agent interacts with simulated quantum optics experiments (see Ref.~\cite{Melnikov2018} and Fig.~\ref{fig:envs} left). 
As observations, the agent receives information about the current experimental setup. 
As actions, the agent sequentially places linear optical elements such as beam splitters, mirrors, Dove prisms, and holograms along the path of a beam that can be described by a quantum state of light. The degree of freedom that is manipulated is the orbital angular momentum (OAM) of photons ~\cite{allen_1992_orbital}, which spans a high-dimensional (in principle infinite) Hilbert space.
The task of the agent is to create high-dimensional entanglement between three of the four photons. Since entanglement is a genuinely quantum phenomenon with no classical analog ~\cite{horodecki_2009_quantum}, designing experimental settings to generate entanglement routinely challenges our classical intuition~\cite{Malik2016,erhard_2018_experimental}.
Entanglement is characterized by the so-called Schmidt-rank vector (SRV)~\cite{huber_2013_structure}: This is a three-dimensional vector $(r_1,r_2,r_3)$ where each $r_i$ quantifies the amount of entanglement between photon $i$ and the others. The reward function in this environment depends on the SRV, a positive reward is given only to those SRVs for which $r_i\geq 2$ for all $i=1,2,3$. We will not distinguish the different orderings of the three SRV components. 
As an initial setup, the agents can operate on two pairs of entangled photons created by a spontaneous parametric down-conversion (SPDC) in two nonlinear crystals~\cite{Malik2016}. The detection of one photon at the end, in combination with the linear optical elements mentioned above, enables the probabilistic generation of entanglement.
Further details on the environment are provided in Appendix~\ref{appendix:qo_env}.

\subsubsection{Intrinsically motivated agent}

The environment described above has a unique starting state corresponding to two pairs of entangled photons in four distinct spatial modes. As the task of standard RL agents is to maximize the future expected (extrinsic) reward as defined in Eq.~(\ref{eq:reward_rl}), we can expect that a trained agent always places elements that reproduce the simplest successful setup. Instead, we ensure continuous exploration by using an intrinsically motivated agent based on a variant of Monte Carlo tree search (MCTS) \cite{Salge2018}. In standard MCTS \cite{Browne2012}, a policy is derived from a search tree built for each state, by keeping track of the rewards and number of state visits during the interaction with a known model of the environment. The intrinsic motivation is added to the MCTS in the form of a novelty count and a boredom mechanism. The novelty count guarantees that each encountered rewarded state is only rewarded once, while the boredom mechanism favors the exploration of less explored action sequences. Further details of the RL algorithm are provided in Appendix~\ref{appendix:rl}. We also analyze the performance of the agent in this environment in Appendix~\ref{appendix:qo_env}.

\subsubsection{Gadget analysis}

Once data has been collected, we analyze the gadgets and clusters found by the data mining and clustering algorithms described in Sec.~\ref{sec:methods:data_mining} and~\ref{sec:methods:clustering}, respectively. 
We separately mined 10 datasets $D_A^{(i)}$ generated by as many independent MCTS agents, labeled by $i \in {1,..,10}$. These datasets have a fixed size of $|D_A^{(i)}|=10000$ action sequences, each with length $T=12$. The mined datasets yield gadgets w.r.t. $F_\textrm{min}^{(i)},C_\textrm{min}^{(i)},I_\textrm{min}^{(i)},G_\textrm{min}=1$ and $D_A^{(i)}$ according to Def.~\ref{def:gadget}. The exact values for these thresholds are listed in Appendix~\ref{appendix:results_qo}, while all gadgets are reported in Appendix~\ref{appendix:qo_env}.\\
To facilitate a direct comparison to Ref.~\cite{Melnikov2018}, we consider only gadgets that are rewarded in a similar way as in \cite{Melnikov2018}, that is, gadgets that produce quantum states with SRVs $(r_1,r_2,r_3)$ such that $r_i\geq 2$ for all $i=1,2,3$. We consider this an additional postprocessing step as described in Sec.~\ref{subsubsec:postprocessing}; all other more generic postprocessing steps are detailed in Appendix~\ref{appendix:results_qo}. 
Once all datasets have been mined, we consider two ways of clustering them: Clustering gadgets by their SRV and by utility, respectively, as described in Sec.~\ref{sec:methods:clustering}. Both methods were applied to the same gadgets.\\

\begin{figure}[t!]
\begin{center}
\includegraphics[width=15cm]{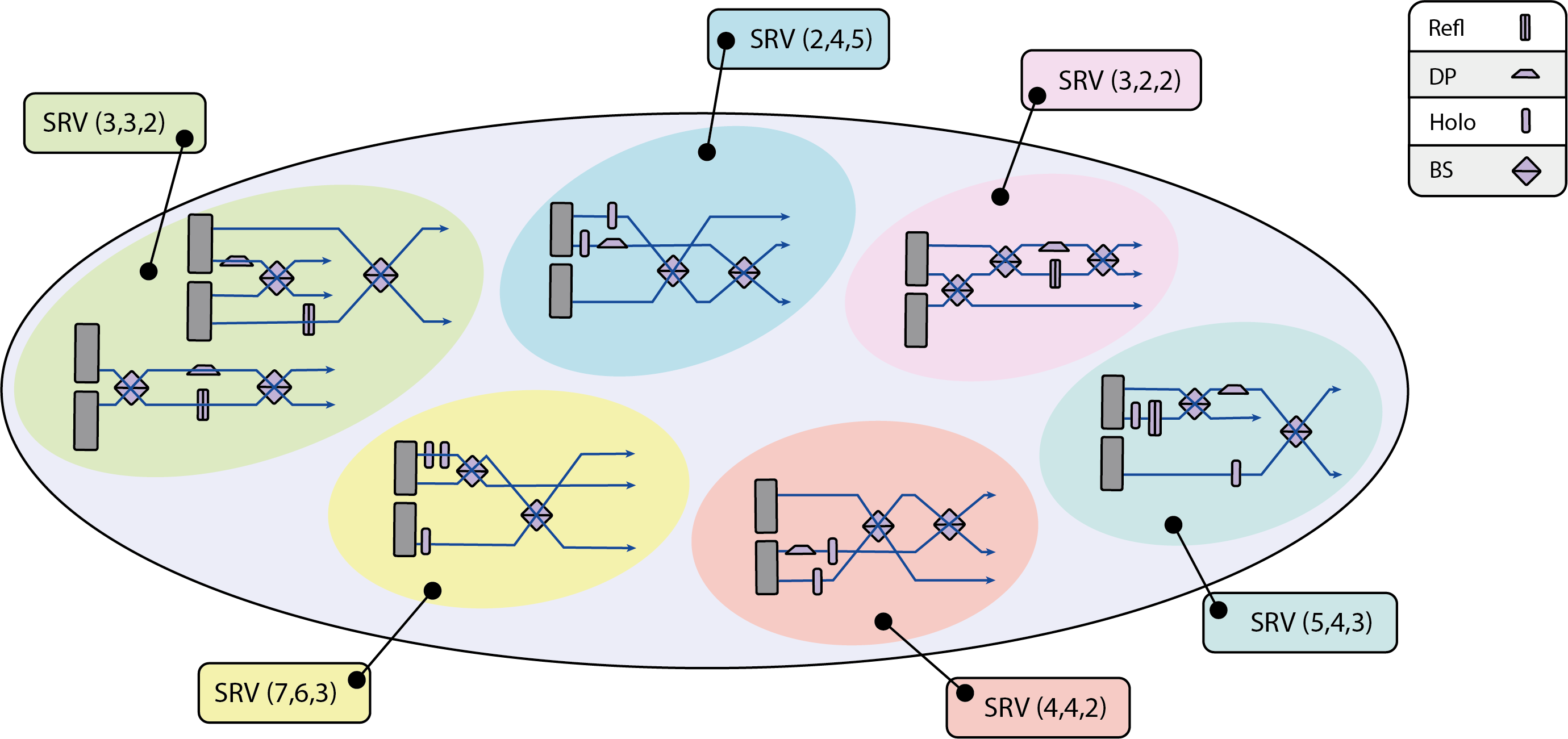}
\caption{\textbf{Quantum optics gadgets clustered by SRV.} Here we show examples of quantum optics gadgets labeled by their SRVs. Gadgets have been mined from datasets generated by 10 intrinsically motivated MCTS agents. We find various interferometers in the clusters. Each of them consists of combination of at most four different types of optical elements: mirrors (Refl), Dove prisms (DP), holograms (Holo), and beam splitter (BS). Most notably, the cluster corresponding to SRV $(3,3,2)$ (green) contains the most important gadgets found in Ref.~\cite{Melnikov2018}, i.e., a Mach-Zehnder interferometer and an equivalent (non-local) version thereof.}
\label{fig:results_qo_SRV}
\end{center}
\end{figure}

\textbf{SRV-based clustering.}  Fig.~\ref{fig:results_qo_SRV} illustrates the results of gadgets clustered by their SRV. We find that the cluster labeled by the SRV $(3,3,2)$ indeed contains the same gadgets retrieved in Ref.~\cite{Melnikov2018}. The other clusters contain additional gadgets that also correspond to interferometric setups. Further details on these clusters are shown in Tab.~\ref{tab:srv_qo}. This clustering method can be used to look up how to generate different SRV states. Utility-based clustering can then be used to gain insights into the composition of the gadgets, as we discuss in the next section. \\

\begin{figure}[t!]
\begin{center}
\includegraphics[width=15cm]{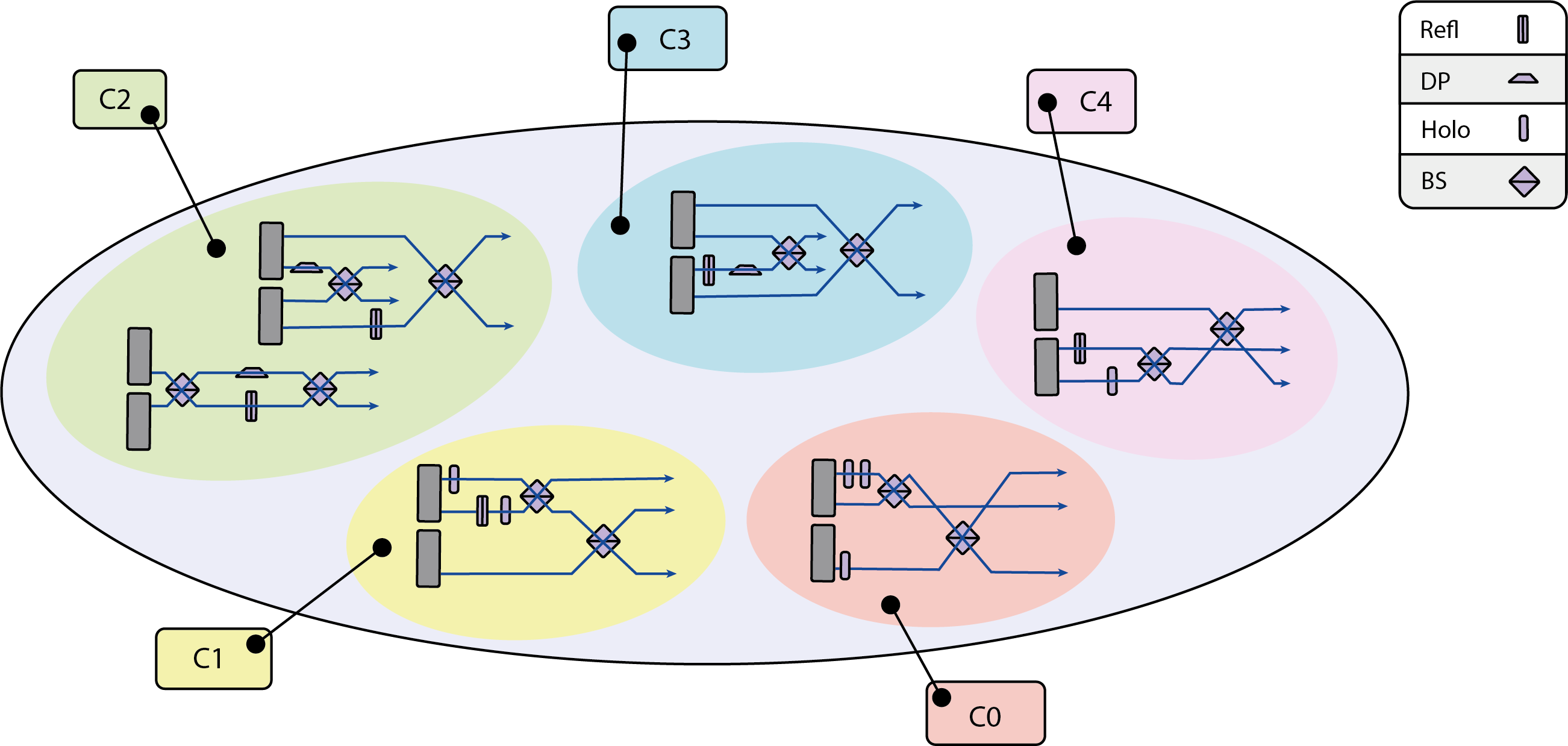}
\caption{\textbf{Quantum optics gadgets clustered by utility.} Here we show examples of quantum optics gadgets labeled by their utility cluster. Gadgets have been mined from datasets generated by a total of 10 intrinsically motivated MCTS agents. The clusters are separated according to the composition of the gadget. Each gadget consists of a combination of at most four different types of optical elements: mirrors (Refl), Dove prisms (DP), holograms (Holo), and beam splitter (BS). Most notably, we find that the two equivalent Mach-Zehnder interferometers (shown in the green C2 cluster) are contained in the same cluster even though they have two different representations. In this environment the $(3,3,2)$-states is most easily generated by the agent, this leads to the majority of the clusters (Clusters C1-C4) containing mostly gadgets to generate this state, while only one cluster(cluster C0) contains the gadgets that generate the less frequent larger SRV states.}
\label{fig:results_qo_utility}
\end{center}
\end{figure}

\textbf{Utility-based clustering.} We illustrate the results of utility-based clustering in Fig.~\ref{fig:results_qo_utility}. For this, we represent each gadget as a four-component vector, where each component counts how many optical elements of one type (beam splitter, hologram, Dove prism, mirror) appear in the gadget. The utility of each component is set to one: $\alpha_i=1 \,\forall i\in\{1,...,4\}$. The distance between two gadgets is the Euclidean distance between their composition encodings. Thereby gadgets are clustered by their composition, yielding insight into the optical elements needed to generate a specific SRV state.  In SRV-based clustering, the cluster associated with SRV (3,3,2) contains the largest number of gadgets. In contrast, utility-based clustering breaks this cluster into five smaller clusters, each corresponding to specific sequences of elements that generate the $(3,3,2)$-state. Cluster C4 mostly contains gadgets that consist of a mirror, a hologram, and two beam splitters. Cluster C2 contains the two equivalent Mach-Zehnder interferometers described in Ref.~\cite{Melnikov2018}: two equivalent arrangements of a mirror, a Dove prism, and two beam splitters. We also note that we find that, compared to cluster C2, the gadgets in cluster C3 often contain one additional optical element: a hologram. Equivalently, C1 contains one extra element, a hologram, w.r.t. C4. An additional hologram is a useful to reach higher SRV-states when combined with an additional beam splitter. Two gadgets in C0, on the other hand, produce a $(3,3,2)$-state with additional holograms and mirrors. There is only a single unclustered element, it is the only gadget that creates a $(3,3,2)$-state with the same elements as C4 with one additional mirror. Thus, C2 and C4 contain the shortest gadgets that produce the $(3,3,2)$-states. Cluster C0 differs from the others, as six out of the eight gadgets produce states where at least one SRV component is greater than three. Interestingly, gadgets that generate these larger SRV states have a similar composition: All of these consist of holograms followed by beam splitters. Such a setup increases the dimensionality of the photons. A single hologram followed by a beam splitter is a gadget that is also mentioned in Ref~\cite{Melnikov2018} and was the result of an extensive post-hoc analysis. Instead, here it arises naturally in a single cluster.
Further details on the clusters are shown in Table~\ref{tab:utility_qo}.

\textbf{Interpretation.} It is difficult to understand the behaviour, i.e., the policy of an exploration-driven agent such as the MCTS algorithm because its behaviour constantly evolves. Here, we provide an alternative way of analyzing its policy by means of clustering gadgets.
As an example, from these gadgets alone we can learn that interferometers are particularly useful in this environment. In addition, the clusters in Figs.~\ref{fig:results_qo_SRV} and~\ref{fig:results_qo_utility} give us a strong starting point to study similarities between gadgets, which could potentially if further automation is applied lead the agent to the discovery of the Klyshko wavefront picture~\cite{klyshko_1988_simple,aspden_2014_experimental}, where the two gadgets displayed in cluster C2 in Fig.~\ref{fig:results_qo_utility} are equivalent.
We provide a detailed list of all gadgets in Appendix~\ref{appendix:results_qo}.

\subsection{Discovering quantum information gadgets}\label{sec:results_qi}

In this section, we use our architecture to discover gadgets useful for quantum information. Quantum information processing functions in a way that is fundamentally different from classical information. For example, measuring quantum information generally entails that the information in that degree of freedom is destroyed. At the same time, quantum information theory allows for fascinating and counter-intuitive phenomena like teleportation and non-locality. Here, we want to explore these phenomena through the lens of gadgets.

\subsubsection{Quantum information environment}
In this RL environment, an agent interacts with a simulated 4-qubit quantum circuit (see Fig.~\ref{fig:envs} right). 
As observation, the agent perceives the current quantum circuit consisting of various gates and measurements.
As actions, the agent can sequentially place quantum gates and measurements, including single-qubit rotations and two-qubit entangling gates, as well as single-qubit measurements and two-qubit Bell measurements.
The task of the agent is ``to keep the quantum information alive'', that is, to avoid losing the information that is initially encoded in the first (top) qubit of the circuit. As the initial setup, the agent is confronted with an incomplete quantum circuit, consisting of random quantum operations only at fixed positions, i.e., layers (marked with an E in Fig.~\ref{fig:envs} right). The agent performs actions until the circuit has been completed. Whenever the circuit preserves the quantum information until the end, the environment issues a positive reward. Conversely, whenever the quantum information is destroyed, through e.g. measurements, a negative reward is given. A more detailed description of this environment and the reward function can be found in Appendix~\ref{appendix:qi_env}.

\subsubsection{Deep RL agent}

The environment described above has so-called random starts, i.e., the initial state is sampled randomly from a finite set of initial states. For an agent to learn in this kind of environment, it needs to be able to adapt to many different situations and develop multiple strategies at the same time. In such environments with large observation spaces, deep RL is particularly advantageous due to its ability to generalize. Generalization is the ability to perform well when confronted with unseen data by identifying meaningful similarities between previously seen and novel (i.e., unseen) data. Here, we apply a double deep Q-network (DDQN)~\cite{VanHasselt2015}. DDQN is a Q-learning algorithm based on the standard deep Q-network (DQN) \cite{Mnih2015}, which features two neural networks to increase the stability of the prediction of Q-values. 
Further details on the RL algorithm are provided in Appendix~\ref{appendix:rl}. We also analyze the performance of the agents in this environment in Sec.~\ref{appendix:qi_env}.

\subsubsection{Gadget analysis}

We analyze the gadgets and clusters found by the data mining and clustering algorithms described in Sec.~\ref{sec:methods:data_mining} and~\ref{sec:methods:clustering}, respectively.
We consider three individual DDQN agents, each of which independently generates a dataset $D_A^{(i)}$ ($i \in \{1,2,3\}$) of unique rewarded circuits, with fixed size $|D_A^{(i)}|=8\cdot 10^4$. The mining algorithm then outputs gadgets w.r.t. $F_\textrm{min}^{(i)},\: C_\textrm{min}^{(i)},\: I_\textrm{min}^{(i)}$, $G_\textrm{min}=1$ and $D_A^{(i)}$ (see Def.~\ref{def:gadget}). The values for the thresholds are automatically adjusted to output between 1 and 10 gadgets per dataset (see Appendix~\ref{appendix:dm}). The final threshold values are given in Appendix~\ref{appendix:results_qi}.

Each operation that the agent can place on the circuit is associated with four features (see also Appendix~\ref{appendix:qi_env}), which describe (i) whether the operation is a gate or a measurement, (ii) whether it is a 1-qubit or 2-qubit operation, (iii) the qubits it is applied on and (iv) the specific gate or measurement. This action encoding provides a minimal structure to the actions without inputting prior knowledge about the environment. Features (i) and (ii) are coarse-grained descriptions of the available operations, and feature (iii) is potentially useful, especially if there is no prior assumption about symmetries in the environment. As explained in Sec.~\ref{subsubsec:postprocessing}, giving this additional structure to the actions allows us to control the granularity of the mining process. Here, we coarse-grain the mining by removing the last feature of each operation (see Appendix~\ref{appendix:results_qi} for further information on gadget mining).

After the mining algorithm outputs gadgets for each dataset $D_A^{(i)}$, we collect them all together and apply utility- and context-based clustering as described in Sec.~\ref{sec:methods:clustering}. Both clustering methods were applied to the same gadgets. A detailed description of the distance measures used for clustering is given in Appendix~\ref{appendix:results_qi}.\\

\textbf{Utility-based clustering.}
We illustrate the result of clustering gadgets by their utility in Fig.~\ref{fig:results_qi_utility}. We consider that each action feature has the same utility. As we describe in Appendix~\ref{appendix:results_qi}, each gadget is encoded into a vector that counts how many 1-qubit and 2-qubit gates and measurements it contains. We define the distance between two gadgets as the euclidean distance between their encodings (see also App.~\ref{appendix:results_qi}), which implies that two gadgets with similar tallies are closer together, even if the order is different and operations do not commute. Indeed, this is what we find in the clusters of Fig.~\ref{fig:results_qi_utility}. 
A list of all gadgets and the corresponding cluster label is given in Table~\ref{tab:utility_qi}. Even though these results do not give us new insight into the environment, in this case, they show that this clustering method is a neat way of grouping gadgets based on the operations they contain, which can be particularly useful when the set of available operations is large. \\

\begin{figure}[t]
\begin{center}
\includegraphics[width=10.6cm]{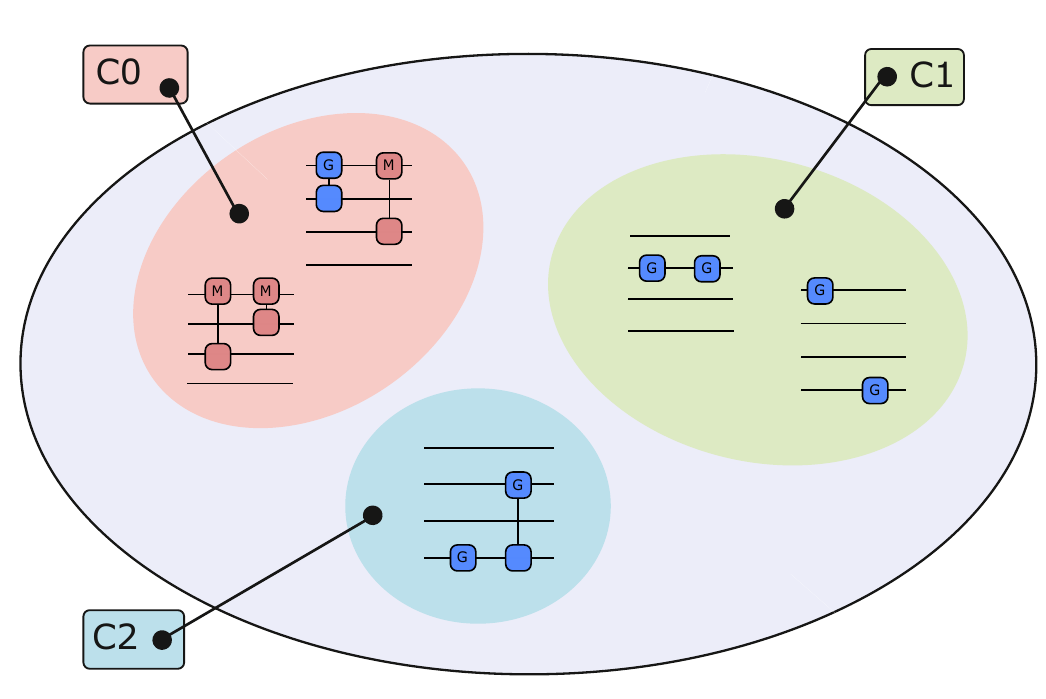}
\caption{\textbf{Quantum information gadgets clustered by utility.} Here we show examples of quantum information gadgets for each cluster. We can clearly distinguish clusters that correspond to gadgets that contain single-qubit gates, multi-qubit gates, multi-qubit measurements, and combinations thereof. The complete list of all gadgets and the corresponding utility cluster is given in Table~\ref{tab:utility_qi} (Appendix~\ref{appendix:results_qi}).}
\label{fig:results_qi_utility}
\end{center}
\end{figure}

\begin{figure}[t]
\begin{center}
\includegraphics[width=12.6cm]{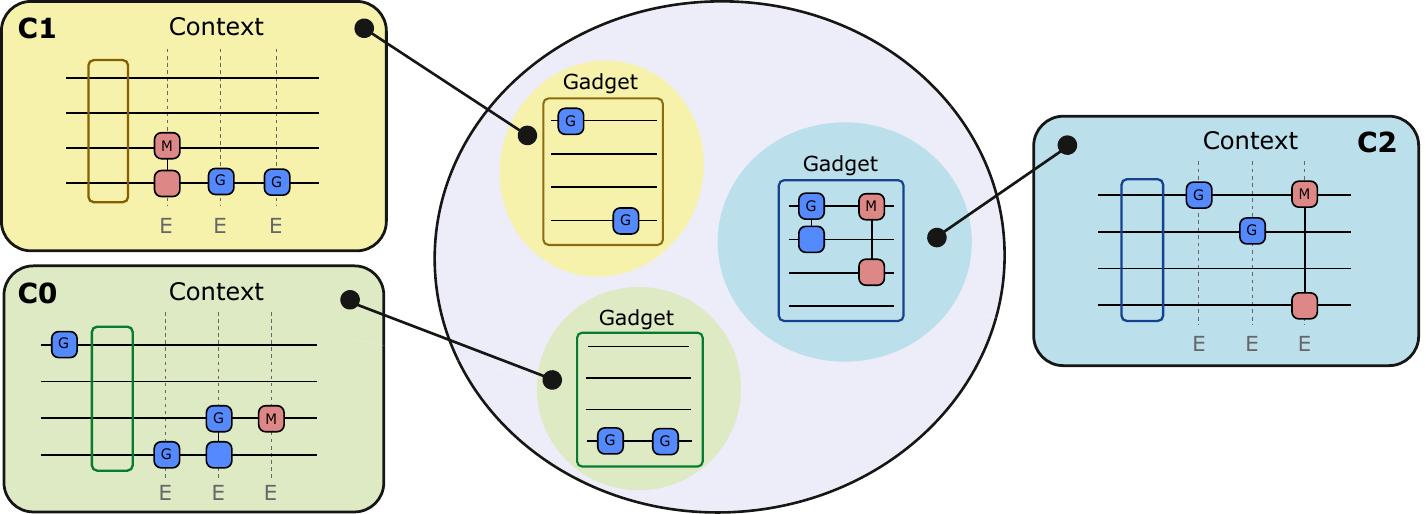}
\caption{\textbf{Quantum information gadgets clustered by context.} In the center, we show examples of quantum information gadgets for each cluster. The cluster labels on the sides show an example of a circuit from the initialization set (context) of the depicted gadget. The empty square in the circuit indicates where the gadget will be placed. Operations marked with an E are randomly chosen by the environment as the initial start. We find that the clustering algorithm distinguishes gadgets that are used in circuits that do not need to be corrected to protect the information (clusters C0 and C1), from teleportation gadgets (cluster C2) in the form of entangling gates between qubits and measurements. Circuits in the initialization sets of gadgets in C2 usually contain a measurement on the first qubit, which destroys the quantum information. To correct these circuits, the agent learned to use the gadgets in C2, which teleport the information to a different qubit to circumvent the measurement. A complete list of all gadgets and their cluster labels can be found in Table~\ref{tab:context_qi} (Appendix~\ref{appendix:results_qi}).}
\label{fig:results_qi_context}
\end{center}
\end{figure}

\textbf{Context-based clustering.}
We illustrate the results of clustering gadgets by their context in Fig.~\ref{fig:results_qi_context}. This clustering method complements and refines the utility-based method as it does not depend on the encoding of the gadget but rather on that of the observation space of the environment. In this metric space, gadgets are closer together if they are used on similar quantum circuits, i.e. if the agent had observed similar states before using them. Gadgets that belong to cluster C1 are placed at the beginning of the circuit when the initial circuit is harmless (it does not contain any measurement that destroys the quantum information). Gadgets in cluster C0 are placed later in the circuit (see example in Fig.~\ref{fig:results_qi_context}). These gadgets are also applied when the circuit does not need to be corrected to protect the information. The gadgets in cluster C2 are used by the agent to correct initial circuits that contain a measurement on the first qubit (see example in Fig.~\ref{fig:results_qi_context}). Interestingly, gadgets in this cluster contain 2-qubit gates and measurements that teleport the quantum information from the first qubit to some other qubit in the circuit to circumvent the measurement initially placed by the environment. Detailed information on these results is given in Table~\ref{tab:context_qi} and further discussed in Appendix~\ref{appendix:results_qi}.\\

\textbf{Interpretation.}
As this environment is inherently random, it is difficult to understand the policy of an RL agent, i.e., why the agent does what it does, especially if it is a black-box neural network. We provide an alternative way of analyzing such a policy by means of gadgets that are then clustered according to several methods. From gadgets alone, we do not learn much as they consist of various combinations of all actions. However, the clustering discloses relevant information about the environment. First, we can distinguish gadgets by the types of operations they use (see Fig.~\ref{fig:results_qi_utility}). Second, we can analyze them by the similarity of the contexts they appear in (see Fig.~\ref{fig:results_qi_context}). The latter approach provides interesting insights into the environment. For example, we observe that entangling gates on the first qubit are used when the context involves a measurement on that qubit. 


\section{Conclusion}

Machine learning has already been acknowledged as part of the modern-day physicist's toolbox~\cite{zdeborova_2017_new}, and it promises to take on a more fundamental role that goes beyond problem-solving. 
One challenge to the development of automated research assistants is the fact that most modern ML approaches operate as black boxes. 
While research assistant systems that solve a complicated problem are already a valuable asset, we want to develop them further and gain insights into the solution process.

In this work, we develop an architecture that operates on top of reinforcement learning agents to autonomously extract promising patterns (gadgets) from the agent's behavior in a post-hoc analysis. In this way, we provide an additional layer of transparent analysis to the agent's policy beyond extracting a single solution to the problem at hand. We demonstrate this method in two quantum physics environments, which are particularly interesting because of their often counter-intuitive nature. We subjected the resulting gadgets and clusters to an expert analysis and confirmed that most of them are indeed meaningful in these environments.

We expect that the framework developed in this work will form the basis of more general architectures that further improve the transparency of RL agents in quantum science and beyond. 
For example, we currently use intrinsically motivated RL agents to stir the discovery of gadgets in simulated quantum optics experiments. In the future, we expect intrinsic motivation to become part of the gadget discovery architecture, such that the information about current gadgets is fed back into the RL agent via an intrinsic reward function. In this way, we can stimulate either the discovery of new gadgets or the use of already known gadgets (i.e., in the spirit of hierarchical learning~\cite{Pateria2021}). This approach can be facilitated and made efficient by the use of well-established data stream pattern mining methods~\cite{Hijawi2015, Xu2009}.
Moreover, while the current architecture focuses on discrete action sequences, in the future this can be extended to continuous action spaces, as well as actions sampled from a policy. Indeed, we observe that gadgets are closely related to so-called \textit{options} \cite{Sutton1999}, a well-established concept in RL. Specifically, options are triples $(\mathcal{I}, \pi, \mathcal{T})$ consisting of a sub-policy $\pi$, an \textit{initialization condition} $\mathcal{I}$ and \textit{termination condition} $\mathcal{T}$. In the context of our architecture, we would consider $\pi$ as a gadget and use $\mathcal{I}$ and $\mathcal{T}$ in the same way that we use the initialization set for gadget clustering now. However, many interesting problem settings involve a discrete action space \cite{Nautrup2018,Ostaszewski2021}, where the presented architecture can already be applied out of the box. 

\section*{Acknowledgements}
We acknowledge support by the Austrian Science Fund (FWF) through the DK-ALM: W1259-N27 and SFB BeyondC F7102, by the Volkswagen Foundation (Az:97721), and the European Research Council (ERC) under Project No. 101055129. This project has received funding from the European Union's Horizon 2020 research and innovation program under the Marie Skłodowska-Curie grant agreement, Grants No. 801110 and No. 885567. It reflects only the author's view, the EU Agency is not responsible for any use that may be made of the information it contains. ESQ has received funding from the Austrian Federal Ministry of Education, Science and Research (BMBWF).


\printbibliography 

\newpage
\appendix
\section{Reinforcement learning}\label{appendix:rl}
Here, we motivate and discuss in more detail the RL algorithms that are used to collect data in Sec.~\ref{sec:results}. Let us consider the two different types of RL environments that appear in Sec.~\ref{sec:results} and how they motivate our choice of RL agents. Specifically, we consider environments with and without random starts. Our goal is to discover gadgets that are representative of the environment or policy. If the environment is deterministic and has a fixed starting state (such as the quantum optics environment in Sec.~\ref{sec:results_qo}), then a normal agent is likely going to converge to a single strategy, which would result in just one or two gadgets. In the case of such an environment, we thus employ intrinsically motivated agents to ensure exploration. In contrast, if we consider so-called random starts environments (such as the quantum information environment in Sec.~\ref{sec:results_qi}), the initial state is randomly sampled at every start of an episode. This already ensures sufficient exploration but necessitates the use of an agent that can generalize over a large set of observations. This is why we use a deep RL agent for this environment. In the following, we will describe both types of RL agents used in this work.

\subsection{Monte Carlo tree search with intrinsic motivation}\label{appendix:MCTS}

In environments with sparse \textit{extrinsic} rewards, methods such as Intrinsic Motivation (IM) can enhance an agent's performance. In this work, we use IM to enhance the exploration by the agents when there is a deterministic single-start RL environment. An IM agent receives intrinsic rewards to drive exploration. For example, agents may be self-driven using a sense of surprise \cite{Barto2013, Achiam2016}, curiosity \cite{Frank2014, Houthooft2016, Pathak2017}, boredom \cite{Bellemare2016, Ostrovski2017,Hangl2020}, need for control \cite{Colas2019, Blaes2019}, or empowerment \cite{Mohamed2015, Gregor2016}.

The explorative version of MCTS described here draws upon three of the four main components of the original algorithm: expansion, backpropagation, and selection. Additionally, we consider changes to these subroutines which we partially adopted from \cite{Salge2018}. In the MCTS algorithm, a decision tree is built from interactions with an environment, each node in the tree corresponds to a state encountered in the environment and each directed edge represents a transition to the next state triggered by an action. The first step of IM MCTS is a multi-step expansion that starts from the root node and iteratively builds up a tree. At each node with depth $d$ if there is an action that has not been taken from the corresponding state, then a new node with depth $d+1$ is added as a child node. This procedure is repeated until the depth is equal to an expansion horizon $T$. The environment we consider is episodic with an episode length of $T$ and a single start, thus every rollout starts from the same root node and ends at a leaf node at depth $T$. 
In order to facilitate explorations, we define an intrinsic reward $r$ that only rewards unique goal-states. Each child $s'$ of node $s$ is uniquely defined by the tuple $(s, a)$ and has an associated novelty count $U(s, a)$, which is the number of unique rewarded goal-states reachable from this node. Additionally, the number of visits to a node is tracked with a visit count $N(s, a)$. 
At the end of each rollout, during backpropagation, the novelty counts $U(s,a)$ and the visit counts $N(s,a)$ for all nodes visited during the rollout are updated: $U(s,a):=U(s,a)+r$ and  $N(s,a):=N(s,a)+1$. 
During the rollout, the selection mechanism is triggered, as soon as a state is encountered for which all actions have been visited during a previous rollout. Then the next action is chosen according to the modified upper confidence bound for trees (UCT):   
\begin{equation}
    \text{UCT}_\text{mod}(s,a)=\frac{U(s,a)}{N(s,a)}+0.001*C\sqrt{\frac{\log(N(s))}{N(s,a)}}
\end{equation}
The parent visit count $N(s)=\sum_{a}N(s,a)$ is given by the sum of visit counts to all children nodes. The first term drives exploitation, while the second term drives exploration, so that the hyperparameter $C$ balances exploration and exploitation. 
The policy for selecting the next action $a$ is greedy: 
\begin{equation}
    a = \argmax_{a'} \text{UCT}_\text{mod}(s,a').
\end{equation}

In the quantum optics environment, an explorative agent is essential, which is why we introduce an additional hyperparameter, a boredom threshold $B$ which resets exploitative agents after some time. The corresponding boredom mechanism resets the novelty count $U(a,s)$ as soon as it reaches the chosen boredom threshold $B$, ensuring further diversity of the collected data set.

\subsection{Double deep Q-network}\label{appendix:DDQN}

Deep RL methods are employed in environments with large state or action spaces where the generalization capabilities of neural networks are beneficial for the agent to improve its performance. At their core, many of these methods are based on adapting the policy of the agent to optimize the return $G_t$: 
\begin{equation}
    G_t= \sum_{k=t+1}^{\infty}\gamma^{k-t-1}r_k,
\end{equation}
with the discount factor $\gamma \in [0,1)$. 
Assuming this figure of merit, each state and action pair $(s,a)$ can be assigned an action-value that quantifies the return expected to be received starting from state $s$ in step $t$ taking action $a$ and subsequently following policy $\pi$. 
\begin{equation}
    q_\pi(s,a)=\mathbb{E}_{\pi}\left[ G_t|s, a \right]
\end{equation}
The aim is to find the optimal policy, i.e., a policy with a greater or equal expected return compared to all other policies for all states. Such a policy can be derived from the optimal action-value function $q_*$. The recursive relationship between the value of the current state and the next state can be expressed in the Bellman optimality equation, which, when solved, yields the optimal policy.
\begin{equation}
   q_*(s,a)= \mathbb{E}\left[r_{t+1}+\max_{a'}q_*(s_{t+1},a')|s, a \right] 
\end{equation}
Instead of directly solving the Bellman optimality equation, in value-based RL, the aim is to learn the optimal action-value function from data samples to derive the optimal policy from the learned values.
One of the most prominent value-based RL algorithms is $Q$-learning, which was designed for discrete state and action spaces, where each state-action pair $(s,a)$ is assigned a so-called $Q$-value $Q(s,a)$ which is updated to approximate $q_*$. Starting from an initial guess for all values $Q(s,a)$, the values are updated for each state action pair $(s,a)$ while the agent interacts with the environment according to the following update rule:   
\begin{equation}
    Q(s,a)\leftarrow Q(s,a)+\alpha \left(r+\gamma \max_{a'} Q(s',a')-Q(s,a)\right),
\end{equation}
where $\alpha$ is the learning rate and $s'$ is the next encountered state after taking action $a$ in state $s$. 
This algorithm has been proven to converge under certain criteria, one of which is a policy that is infinitely exploring but greedy in the limit. 
In practice, a popular choice for a sufficiently explorative policy is the $\epsilon$-greedy policy: 
\begin{equation}
 \pi(a|s)= \begin{cases}
1-\epsilon_t \quad \text{for}\quad a=\max_{a'}Q(s,a')  \\
\epsilon_t \quad \text{otherwise} \\
\end{cases}  
\end{equation}

In the following, we will look at how $Q$-learning can be extended to large state and action spaces through the use of neural networks (NNs) as function approximators. 

 However, in practice, neural networks  work exceedingly well, solving tasks that could not be handled by prior RL methods. 
An essential requirement for NN training, independently and identically distributed data, is not naturally provided by the sequential nature of RL data. This problem is circumvented by experience replay. This method trains the neural network with batches of experiences previously divided into single episode updates and randomly sampled from a memory.
To stabilize training, two NNs are employed, a policy network, that is continuously updated and a target network that is an earlier copy of the policy network. The policy network is used to estimate the current value while the target network is used to provide a more static target value $Y$:
\begin{equation}
Y_\text{DQN}= r+\gamma \max_{a'} Q_\text{target}(s',a')
\end{equation}
Using the Double Deep Q-Nettwork (DDQN) algorithm, the action for the target value is sampled from the policy network to reduce the overestimation bias present in standard DQN. The corresponding target is defined by: 
\begin{equation}
  Y_\text{DDQN}= r+\gamma Q_\text{target}(s', \argmax_{a'} Q_\text{policy}(s',a')). 
\end{equation}
This target value will be approximated using a chosen loss function, we choose a smooth L1-Norm loss. 

\section{Data mining algorithms}\label{appendix:dm}
Here, we describe an efficient algorithm for identifying interesting subsequences, based on the SPADE algorithm \cite{Zaki2001} to find frequent subsequences. The algorithm starts by collecting all length-one subsequences with $I(\tilde{d})>I_{\min}$ into a set of interesting subsequences $\tilde{D}$. The minimum interestingness $I_{\min}$ is a user-defined threshold. All length-one subsequences in this set are then combined into length-two subsequences. Every length-two subsequence with $F(\tilde{d})>F_{\min}$, $C(\tilde{d})>C_{\min}$ and $I(\tilde{d})>I_{\min}$ will be added to the set of interesting subsequences $\tilde{D}$. The minimum cohesion $C_{\min}$ and the minimum support $F_{\min}$ are user-defined thresholds. The set of interesting subsequences $\tilde{D}$ will be iteratively filled with longer subsequences until a predefined maximum subsequence length is reached.  

After the generation of the most frequent and compact subsequences, the output patterns may be filtered in a postprocessing step. For our purposes, we filtered by length and coverage. 
The length filter discards all patterns that are shorter or longer than a pre-defined interval. The coverage filter ensures that the maximum number of patterns that appear in each sequence is limited by the maximum coverage value. This filter works as follows: for each pattern with more than one element, starting from the patterns with the highest interestingness\footnote{In case of equal interestingness, longer patterns have preference.}, it checks in which sequences the pattern appears, or, in other words, which sequences are ``covered" by the pattern. Each sequence can only be covered by a maximum value of patterns, set by the user. Patterns that are left with no sequence to cover are discarded. Hence, choosing a low maximum coverage value means that more patterns will be filtered out. Additionally, we define an RL data-specific mechanism that filters patterns by their reward.

We added an hyperparametrization scheme to the mining procedure that automatically adjusts the threshold values $F_{\min}$, $C_{\min}$, $I_{\min}$ to output a number of patterns that lies within an interval. Therefore, the user can either set fixed values for the thresholds, or choose an interval for the number of output patterns instead, to ensure that the algorithm does not yield too few nor too many patterns. 

The algorithm described above is straightforwardly extended to find gadgets by restricting the underlying dataset to a set of action sequences of interaction sequences $d$ that have a return $G(d)\geq G_\textrm{min}$ as described in Sec.~\ref{sec:methods:RL}.

\section{Hierarchical density-based spatial clustering of applications with noise}\label{appendix:HDBSCAN}
To cluster the gadgets we use the hierarchical density-based spatial clustering of applications with noise (HDBSCAN) algorithm~\cite{Campello2013}, which is a hierarchical clustering algorithm that extracts a set of clusters based on their stability. 
This is done in a series of steps which we will briefly review in this section. 
First, the data will be transformed such that dense regions stay dense and sparse regions become even sparser. This is achieved by transforming the metric space using the so-called core distance $\text{core}_k(x)$, which is the distance within which the next $k$ neighboring data points can be reached. Then, the mutual reachability distance, which either preserves the original distance or increases it to the larger core distance of two data points, can be defined: 
\begin{equation}
    d_\text{reach-k}(a,b)=\max\{\text{core}_k(a), \text{core}_k(b), d(a,b)\}.
    \end{equation}
Here, $d(a,b)$ is the distance between the data points $a$ and $b$ under the original distance metric. 
 
 Then, the data is represented in a graph, where each node represents a data point and the edges between the nodes are weighted by the mutual reachability distance. If all edges above a certain mutual reachability distance threshold are removed, the graph breaks into subgraphs. Then, a hierarchy is established that orders the graph from completely connected to disconnected under the varying distance threshold. Such a hierarchy can be extracted from the minimum spanning tree, a loop-free tree that contains all $M$ nodes of the tree and $M-1$ edges with the lowest sum of weights given all the original weighted edges. Prim's algorithm is used to obtain this graph.  

After building the hierarchy from progressively more disconnected subgraphs, the hierarchy can be further condensed by introducing a minimal cluster size.The clusters are retrieved from the hierarchy using the following definition of the stability of a cluster: 
\begin{equation}
\sum_{p\in C}(\lambda_p-\lambda_\text{birth}), 
\end{equation}
where $C$ is a cluster and $\lambda_p$ is the time when the data point $p$ is removed from the cluster, while $\lambda_\text{birth}$ is the time when this cluster was formed. The clusters are then determined starting from the leaf nodes and declaring all of the leaf clusters as selected. If the sum of the stabilities of the children is higher than the stability of the parent, then the stability of the parent is set to the sum of its children's stabilities. If the sum of stabilities is lower, the parent will be declared selected instead of its children. When the root node is reached the algorithm terminates and outputs the set of selected clusters. Further details on the algorithm can be found in {\href{https://hdbscan.readthedocs.io/en/latest/how_hdbscan_works.htm}{this tutorial}}.

\section{Further results}
In this section, we describe the environments and the obtained results in more detail. Specifically, we list all gadgets that were found and analyze the different clusters obtained by the clustering.

\subsection{Quantum optics environment}\label{appendix:qo_env}
Here, we describe in detail the quantum optics environment from Ref.~\cite{Melnikov2018} as it is used in Sec.~\ref{sec:results_qo}.
In this environment, the agent is tasked to generate high-dimensional multipartite entangled states. At each time step, the agent interacts with a simulation of an optical experiment placing optical elements in a four-photon path.
At the beginning of each episode, the state is initialized to a product of two OAM-entangled photon pairs, 
\begin{equation}
 \ket{\psi(0)}=\frac{1}{3}\left(\sum_{m=-1}^{1} \ket{m}_a \ket{-m}_b\right) \otimes \left(\sum_{m=-1}^{1} \ket{m}_c \ket{-m}_d\right),
\end{equation}
where indices $a,b,c,d$ specify the four available beam paths of the setup. 
The set of actions, i.e., optical elements, and their unitary effect on the photon state are summarized in Table \ref{table:operations_qo}.
The observation the agent receives is the current setup, i.e., the list of elements placed in the beam paths.  
In every episode, the agent can place 12 elements before the episode terminates. At the end of each episode, the state is measured in path $d$ and the Schmidt-rank vectors (SRV) of the resulting conditional three-photon states are evaluated. The SRV is a real vector containing the rank of the reduced density matrix of each subsystem. That is, considering the tripartite system after measurement, the vector takes the shape $d_{\phi} = (r_1^{\phi},r_2^{\phi},r_3^{\phi})$, where $r_i=\mathrm{rank}(\rho_i)$ for the reduced density matrix $\rho_i=\mathrm{tr}_{\bar{i}}(\rho)$ where $\bar{i}$ is the complement of $i\in\{a,b,c\}$. The environment issues a reward of one if $r_i\geq 2$ for all $i=1,2,3$, otherwise the reward is chosen to be zero. There is one exception to this rule, since the $(4,3,3)$-state is easily generated, this state is not rewarded to facilitate that the agent finds a variety of states.

\begin{table}[]
\centering
\begin{tabular}{l c}
\textbf{Optical Element}  & \textbf{Unitary transformation} \\ \hline
      BS$_{a,b}$: Nonpolarizing symmetric  & $\ket{m}_a \rightarrow (i\ket{-m}_a+\ket{m}_b)/\sqrt{2}$ \\ \vspace{0.2cm}
      50/50 beam splitter & $\ket{m}_b \rightarrow (i\ket{m}_a+i\ket{-m}_b)/\sqrt{2}$ \\ \vspace{0.2cm}
 Holo$_{a(k)}$: Hologram & $\ket{m}_a \rightarrow \ket{m+k}_a$ \\ \vspace{0.2cm}
 DP$_a$: Dove prism & $\ket{m}_a \rightarrow ie^{i\pi m}\ket{-m}_a$\\ 
 Refl$_a$: Mirror & $\ket{m}_a \rightarrow \ket{-m}_a$\\ \hline

\end{tabular}

\caption{Optical elements (actions) and the corresponding unitary transformation on the state space of the quantum optics environment. $m$ is the OAM quantum number of a photon. Subscripts $a,b$ refer to the beam paths. The set of all actions contains the optical elements listed in this table for all combinations of paths $a,b,c,d$ and Holograms with shifts $k \in {\pm1,\pm2}$.} \label{table:operations_qo}
\end{table}


In this environment, the dataset $D$ consists of unique rewarded sequences collected during exploration. Gadgets are retrieved from the dataset $D$ by further processing the data set with the sequence mining algorithm described in Sec.~\ref{sec:methods:data_mining}. The obtained gadgets are listed and analyzed in the following section.

\subsection{Quantum optics gadgets}\label{appendix:results_qo}

\textbf{Gadget mining.}
The dataset $D$ of unique, rewarded action sequences collected during exploration is mined by the sequence mining algorithm (see Sec.~\ref{sec:methods:data_mining} and Appendix~\ref{appendix:dm}). Each sequence consists of 12 optical elements.
Instead of fixing the threshold values for the minimum support $F_\textrm{min}$, minimum cohesion $C_\textrm{min}$ and minimum interestingness $I_\textrm{min}$, the algorithm is set to output between 1 and 5 gadgets for each of the ten agents run in parallel. The thresholds are then automatically adjusted until the correct number of gadgets is retrieved. The resulting threshold values for each agent are given in Table~\ref{table:final_thresholds_qo}. When selecting gadgets for this environment, we apply three additional filters (see Sec.~\ref{appendix:dm} for details): (1) Gadgets are filtered by length to allow only gadgets of length $4-6$. (2) Gadgets are filtered by coverage to ensure that at most 2 gadgets can cover each sequence. (3) We filter gadgets by their reward, which means that only those gadgets are kept that themselves constitute successful experiments. The latter filter is in line with the considerations made in Ref~\cite{Melnikov2018}.
We collect the gadgets from 10 different agents and cluster them in two different ways as described in the following.

\begin{table}[htbp]
\centering
\begin{tabular}{|c|l|l|l|l|l|l|l|l|l|l|}
\hline
\textbf{Agent id} & \multicolumn{1}{c|}{0}    & \multicolumn{1}{c|}{1}    & \multicolumn{1}{c|}{2}    & \multicolumn{1}{c|}{3} & \multicolumn{1}{c|}{4}    & \multicolumn{1}{c|}{5}    & \multicolumn{1}{c|}{6}    & \multicolumn{1}{c|}{7}   & \multicolumn{1}{c|}{8}    & \multicolumn{1}{c|}{9} \\ \hline
\textbf{$F_\textrm{min}$}     & 0.11  & 0.095     & 0.77       & 0.07   &  0.105           & 0.1           &  0.07         & 0.09   &  0.115 & 0.105        \\ \hline
\textbf{$C_\textrm{min}$}      & 0.77     & 0.74        & 0.0847    & 0.7   &  0.76       &  1.02       & 0.7      & 0.73  & 0.78   &    0.76           \\ \hline
\end{tabular}
\caption{Optimized threshold values for the support $F$ and cohesion $C$ for each agent in the gadget mining. The algorithm is set to output between 1-5 gadgets, so that the threshold values are automatically adjusted accordingly. The minimum interestingness $I_\textrm{min}$ is the product of minimum support $F_\textrm{min}$ and minimum cohesion $C_\textrm{min}$.}
\label{table:final_thresholds_qo}
\end{table}

\textbf{Gadget clustering.} In order to help us interpret the list of gadgets obtained from 10 different agents, we cluster them by (i) information provided by the environment in form of the SRV and (ii) the type of operations they contain (see Sec.~\ref{subsec:utility} for a description of utility-based clustering). \\

\textit{SRV-based clustering. }Gadgets are sorted into clusters by the SRV that the gadget would yield if seen as a full experiment. All gadgets, their interestingness, and their clusters are listed in Table~\ref{tab:srv_qo}. \\

\textit{Utility-based clustering. }The clustering is done by the algorithm HDBSCAN (see Appendix.~\ref{appendix:HDBSCAN}), after encoding the gadgets using action features. In RL environments, one can assume knowledge about the structure of the action space. Typically, this is used to encode action by their features. In this case, we describe each action only by one of its features, in particular, the type of optical elements. Thus, this yields a coarse-grain description of the gadgets, which we use to define a distance between them. As described in Section \ref{sec:results_qo}, the four types of optical element are BS, DP, Refl, and Holo. Each gadget is then encoded as a vector of the form $\phi(g) = (n_\text{BS}, n_\text{DP}, n_\text{Refl}, n_\text{Holo})$, where each component is the total number of occurrences of one element type within the gadget. Additionally, we define a utility vector $\alpha = (\alpha_{BS}, \alpha_{DP}, \alpha_{Refl}, \alpha_{Holo})$ that assigns a weight to each optical element type. Since we do not assume any of the elements to be more costly than others, we weight all four types equally, i.e. $\alpha = (1,1,1,1)$. The distance between two gadgets $M(g_1, g_2)$ is then defined as the euclidean distance between $\phi(g_1)$ and $\phi(g_2)$. Clustering gadgets with this distance, we obtain the clusters in Table~\ref{tab:utility_qo}. We observe that gadgets with similar compositions are clustered together. For example, cluster C2 contains only gadgets consisting of two BS, one SP and one Refl. All gadgets, their interestingness, and their clusters are listed in Table \ref{tab:utility_qo}.

\begin{table}
\begin{filecontents*}{Tables/srv_qo_31.csv}
Pattern;SRV; Interestingness;Support;Cohesion;Agent
BS$_{a,d}$,Refl$_{c}$,DP$_{c,1}$,BS$_{b,c}$;$(3,3,2)$;0.1001;0.1001;1.0;8
Holo$_{d(2)}$,Holo$_{b(1)}$,BS$_{a,d}$,Holo$_{c(-2)}$,BS$_{c,d}$;$(5,4,2)$;0.1;0.1001;0.9988;7
DP$_{c,1}$,Holo$_{d(-1)}$,Holo$_{c(-1)}$,BS$_{a,d}$,BS$_{a,c}$;$(4,4,2)$;0.1;0.1;1.0;6
DP$_{a,1}$,BS$_{a,d}$,Refl$_{c}$,BS$_{b,c}$;$(3,3,2)$;0.1;0.1;1.0;0
DP$_{a,1}$,BS$_{a,d}$,Refl$_{b}$,BS$_{b,c}$;$(3,3,2)$;0.1;0.1001;0.999;1
Holo$_{a(-2)}$,Refl$_{b}$,Holo$_{b(1)}$,BS$_{a,b}$,BS$_{b,d}$;$(3,3,2)$;0.1;0.1;1.0;7
Holo$_{d(1)}$,Refl$_{d}$,Holo$_{b(2)}$,BS$_{b,d}$,Refl$_{d}$,BS$_{a,b}$;$(4,4,2)$;0.0999;0.0999;1.0;1
Holo$_{b(2)}$,Refl$_{b}$,BS$_{a,b}$,Holo$_{d(2)}$,DP$_{a,1}$,BS$_{a,d}$;$(5,4,3)$;0.0998;0.0998;1.0;3
Holo$_{b(2)}$,Refl$_{c}$,Holo$_{a(-2)}$,BS$_{a,d}$,DP$_{d,1}$,BS$_{b,d}$;$(5,4,2)$;0.0998;0.0998;1.0;6
Holo$_{b(-2)}$,DP$_{b,1}$,Holo$_{a(2)}$,BS$_{a,d}$,BS$_{b,d}$;$(5,4,2)$;0.0997;0.0999;0.9982;1
Holo$_{a(2)}$,Holo$_{d(2)}$,Holo$_{a(1)}$,BS$_{a,b}$,BS$_{b,d}$;$(7,6,3)$;0.0996;0.0996;1.0;1
BS$_{b,c}$,DP$_{b,1}$,Refl$_{c}$,BS$_{b,c}$;$(3,3,2)$;0.0996;0.0997;0.9987;7
Refl$_{a}$,Holo$_{b(1)}$,BS$_{a,b}$,Holo$_{a(-1)}$,BS$_{b,d}$;$(3,3,2)$;0.0992;0.0994;0.9984;9
Refl$_{c}$,Holo$_{d(-1)}$,BS$_{c,d}$,BS$_{b,d}$;$(3,3,2)$;0.0991;0.0996;0.9948;2
BS$_{a,d}$,Refl$_{c}$,DP$_{c,1}$,BS$_{b,c}$,Holo$_{d(1)}$;$(3,3,2)$;0.0991;0.0991;1.0;8
BS$_{b,d}$,BS$_{a,b}$,DP$_{b,1}$,Refl$_{b}$,BS$_{a,b}$;$(3,2,2)$;0.0989;0.099;0.9988;2
Holo$_{d(1)}$,Refl$_{c}$,BS$_{c,d}$,Refl$_{b}$,BS$_{b,d}$;$(3,3,2)$;0.0983;0.0986;0.9972;5
DP$_{a,1}$,BS$_{a,d}$,Refl$_{b}$,BS$_{b,c}$,Holo$_{c(1)}$;$(3,3,2)$;0.0978;0.0983;0.9953;1
DP$_{a,1}$,BS$_{a,d}$,Refl$_{c}$,BS$_{b,c}$,Holo$_{d(-2)}$;$(3,3,2)$;0.0972;0.0972;1.0;0
Refl$_{d}$,Holo$_{d(-1)}$,BS$_{c,d}$,DP$_{b,1}$,BS$_{b,d}$;$(3,3,2)$;0.0971;0.0975;0.9955;4
Refl$_{c}$,Holo$_{d(-1)}$,BS$_{c,d}$,BS$_{b,d}$,Holo$_{d(1)}$;$(3,3,2)$;0.0951;0.0955;0.9956;2
BS$_{b,c}$,DP$_{b,1}$,Refl$_{c}$,BS$_{b,c}$,Holo$_{d(2)}$;$(3,3,2)$;0.0882;0.0903;0.9773;7
Holo$_{a(-2)}$,Holo$_{d(1)}$,BS$_{a,d}$,Holo$_{c(-2)}$,BS$_{a,b}$;$(4,4,2)$;0.0833;0.1;0.8329;3
Holo$_{a(2)}$,Refl$_{a}$,Holo$_{a(1)}$,BS$_{a,b}$,Holo$_{c(-2)}$,BS$_{a,d}$;$(3,3,2)$;0.083;0.0972;0.8543;8
Holo$_{a(1)}$,Holo$_{d(-1)}$,Refl$_{c}$,BS$_{c,d}$,BS$_{b,d}$;$(3,3,2)$;0.0827;0.0997;0.8296;9
Holo$_{a(-1)}$,Holo$_{c(-2)}$,Refl$_{b}$,BS$_{a,b}$,BS$_{b,d}$;$(3,3,2)$;0.0824;0.0994;0.8293;4
Refl$_{a}$,Holo$_{b(1)}$,BS$_{a,b}$,BS$_{b,d}$,Refl$_{d}$;$(3,3,2)$;0.0822;0.0989;0.831;9
Refl$_{d}$,Holo$_{d(-1)}$,BS$_{c,d}$,BS$_{b,d}$,Holo$_{d(-2)}$;$(3,3,2)$;0.0796;0.0958;0.8313;4
Refl$_{a}$,Holo$_{b(1)}$,BS$_{a,b}$,BS$_{b,d}$;$(3,3,2)$;0.0796;0.0997;0.7981;9
BS$_{a,d}$,Refl$_{c}$,DP$_{c,1}$,BS$_{b,c}$,Holo$_{a(2)}$;$(3,3,2)$;0.0793;0.0959;0.8267;8
Holo$_{d(1)}$,Refl$_{c}$,BS$_{c,d}$,BS$_{b,d}$;$(3,3,2)$;0.0789;0.0992;0.795;5
Refl$_{d}$,Holo$_{d(-1)}$,BS$_{c,d}$,BS$_{b,d}$;$(3,3,2)$;0.0778;0.0979;0.7951;4
DP$_{a,1}$,BS$_{a,d}$,Refl$_{c}$,BS$_{b,c}$,Holo$_{a(2)}$;$(3,3,2)$;0.0757;0.0918;0.8245;0
Refl$_{c}$,Holo$_{d(-1)}$,BS$_{c,d}$,BS$_{b,d}$,DP$_{c,1}$;$(3,3,2)$;0.0756;0.0911;0.8301;2
DP$_{c,1}$,Refl$_{b}$,Holo$_{b(-1)}$,BS$_{a,b}$,BS$_{a,d}$;$(3,3,2)$;0.0707;0.0993;0.7117;7
Holo$_{a(2)}$,Refl$_{a}$,Holo$_{a(1)}$,BS$_{a,b}$,BS$_{a,d}$;$(3,3,2)$;0.07;0.0987;0.7097;8
DP$_{b,1}$,BS$_{b,c}$,Refl$_{d}$,BS$_{a,d}$;$(3,3,2)$;0.0666;0.1;0.6658;3
Refl$_{c}$,BS$_{a,d}$,DP$_{c,1}$,BS$_{b,c}$;$(3,3,2)$;0.0664;0.0996;0.6666;3
DP$_{a,1}$,BS$_{a,d}$,Refl$_{c}$,BS$_{b,c}$,Refl$_{a}$;$(3,3,2)$;0.0628;0.0889;0.7061;0
Refl$_{c}$,Holo$_{d(-1)}$,BS$_{c,d}$,BS$_{b,d}$,DP$_{b,1}$;$(3,3,2)$;0.0623;0.0873;0.7131;2
BS$_{b,c}$,DP$_{b,1}$,Refl$_{d}$,Refl$_{b}$,BS$_{b,c}$;$(3,3,2)$;0.0613;0.0981;0.6248;6
Holo$_{d(1)}$,Refl$_{c}$,BS$_{c,d}$,BS$_{b,d}$,DP$_{d,1}$;$(3,3,2)$;0.0608;0.086;0.7068;5
Holo$_{c(1)}$,Refl$_{c}$,BS$_{c,d}$,BS$_{b,d}$;$(3,3,2)$;0.0565;0.099;0.571;6
Refl$_{d}$,Holo$_{d(-1)}$,BS$_{c,d}$,BS$_{b,d}$,DP$_{d,1}$;$(3,3,2)$;0.049;0.0884;0.554;4
\end{filecontents*}
\pgfplotstabletypeset[font=\small,
    col sep=semicolon,
    string type,
    every row no 1/.style={before row=\hline},
    every row no 5/.style={before row=\hline},
    every row no 8/.style={before row=\hline},
    every row no 2/.style={before row=\hline},
    every row no 43/.style={before row=\hline},
    columns/pattern/.style={column name=Pattern, column type={|c|}},
    columns/interestingness/.style={column name=Interestingness, column type={|c|}},
    columns/support/.style={column name=Support, column type={|c|}},
    columns/cohesion/.style={column name=Cohesion, column type={|c|}},
    columns/info/.style={column name=SRV, column type={|c|}},
    every head row/.style={before row=\hline,after row=\hline},
    every last row/.style={after row=\hline},
    ]{Tables/srv_qo_31.csv}
    
\caption{List of gadgets mined from different datasets collected by ten agents clustered by SRV. The column labeled ``Agent" shows from which agent's dataset the gadget was retrieved. The SRV corresponding to each gadget is listed in the last column. The clusters are separated by horizontal lines. In each cluster, the gadgets are sorted from highest to lowest interestingness value. Additionally, the support and cohesion, which when multiplied together yield the interestingness, are displayed.}
\label{tab:srv_qo}
\end{table}

\begin{table}
\centering
\begin{filecontents*}{Tables/utility_qo_31_wo.csv}
Gadget;Cluster;Prob.;Interestingness;SRV 
DP$_{c,1}$,Holo$_{d(-1)}$,Holo$_{c(-1)}$,BS$_{a,d}$,BS$_{a,c}$;C0;1.0;0.1;$(4,4,2)$
Holo$_{d(2)}$,Holo$_{b(1)}$,BS$_{a,d}$,Holo$_{c(-2)}$,BS$_{c,d}$;C0;1.0;0.1;$(5,4,2)$
Holo$_{b(2)}$,Refl$_{c}$,Holo$_{a(-2)}$,BS$_{a,d}$,DP$_{d,1}$,BS$_{b,d}$;C0;1.0;0.0998;$(5,4,2)$
Holo$_{a(2)}$,Holo$_{d(2)}$,Holo$_{a(1)}$,BS$_{a,b}$,BS$_{b,d}$;C0;1.0;0.0996;$(7,6,3)$
Holo$_{d(1)}$,Refl$_{c}$,BS$_{c,d}$,Refl$_{b}$,BS$_{b,d}$;C0;1.0;0.0983;$(3,3,2)$
Holo$_{a(-2)}$,Holo$_{d(1)}$,BS$_{a,d}$,Holo$_{c(-2)}$,BS$_{a,b}$;C0;1.0;0.0833;$(4,4,2)$
Holo$_{a(2)}$,Refl$_{a}$,Holo$_{a(1)}$,BS$_{a,b}$,Holo$_{c(-2)}$,BS$_{a,d}$;C0;1.0;0.083;$(3,3,2)$
BS$_{b,c}$,DP$_{b,1}$,Refl$_{d}$,Refl$_{b}$,BS$_{b,c}$;C0;1.0;0.0613;$(3,3,2)$
Holo$_{a(-2)}$,Refl$_{b}$,Holo$_{b(1)}$,BS$_{a,b}$,BS$_{b,d}$;C1;1.0;0.1;$(3,3,2)$
Holo$_{d(1)}$,Refl$_{d}$,Holo$_{b(2)}$,BS$_{b,d}$,Refl$_{d}$,BS$_{a,b}$;C1;0.0;0.0999;$(4,4,2)$
Holo$_{b(2)}$,Refl$_{b}$,BS$_{a,b}$,Holo$_{d(2)}$,DP$_{a,1}$,BS$_{a,d}$;C1;0.0;0.0998;$(5,4,3)$
Holo$_{b(-2)}$,DP$_{b,1}$,Holo$_{a(2)}$,BS$_{a,d}$,BS$_{b,d}$;C1;0.0;0.0997;$(5,4,2)$
Refl$_{a}$,Holo$_{b(1)}$,BS$_{a,b}$,Holo$_{a(-1)}$,BS$_{b,d}$;C1;1.0;0.0992;$(3,3,2)$
Refl$_{c}$,Holo$_{d(-1)}$,BS$_{c,d}$,BS$_{b,d}$,Holo$_{d(1)}$;C1;1.0;0.0951;$(3,3,2)$
Holo$_{a(1)}$,Holo$_{d(-1)}$,Refl$_{c}$,BS$_{c,d}$,BS$_{b,d}$;C1;1.0;0.0827;$(3,3,2)$
Holo$_{a(-1)}$,Holo$_{c(-2)}$,Refl$_{b}$,BS$_{a,b}$,BS$_{b,d}$;C1;1.0;0.0824;$(3,3,2)$
Refl$_{d}$,Holo$_{d(-1)}$,BS$_{c,d}$,BS$_{b,d}$,Holo$_{d(-2)}$;C1;1.0;0.0796;$(3,3,2)$
Holo$_{a(2)}$,Refl$_{a}$,Holo$_{a(1)}$,BS$_{a,b}$,BS$_{a,d}$;C1;1.0;0.07;$(3,3,2)$
BS$_{a,d}$,Refl$_{c}$,DP$_{c,1}$,BS$_{b,c}$;C2;1.0;0.1001;$(3,3,2)$
DP$_{a,1}$,BS$_{a,d}$,Refl$_{c}$,BS$_{b,c}$;C2;1.0;0.1;$(3,3,2)$
DP$_{a,1}$,BS$_{a,d}$,Refl$_{b}$,BS$_{b,c}$;C2;1.0;0.1;$(3,3,2)$
BS$_{b,c}$,DP$_{b,1}$,Refl$_{c}$,BS$_{b,c}$;C2;1.0;0.0996;$(3,3,2)$
DP$_{b,1}$,BS$_{b,c}$,Refl$_{d}$,BS$_{a,d}$;C2;1.0;0.0666;$(3,3,2)$
Refl$_{c}$,BS$_{a,d}$,DP$_{c,1}$,BS$_{b,c}$;C2;1.0;0.0664;$(3,3,2)$
BS$_{a,d}$,Refl$_{c}$,DP$_{c,1}$,BS$_{b,c}$,Holo$_{d(1)}$;C3;1.0;0.0991;$(3,3,2)$
DP$_{a,1}$,BS$_{a,d}$,Refl$_{b}$,BS$_{b,c}$,Holo$_{c(1)}$;C3;1.0;0.0978;$(3,3,2)$
DP$_{a,1}$,BS$_{a,d}$,Refl$_{c}$,BS$_{b,c}$,Holo$_{d(-2)}$;C3;1.0;0.0972;$(3,3,2)$
Refl$_{d}$,Holo$_{d(-1)}$,BS$_{c,d}$,DP$_{b,1}$,BS$_{b,d}$;C3;1.0;0.0971;$(3,3,2)$
BS$_{b,c}$,DP$_{b,1}$,Refl$_{c}$,BS$_{b,c}$,Holo$_{d(2)}$;C3;1.0;0.0882;$(3,3,2)$
BS$_{a,d}$,Refl$_{c}$,DP$_{c,1}$,BS$_{b,c}$,Holo$_{a(2)}$;C3;1.0;0.0793;$(3,3,2)$
DP$_{a,1}$,BS$_{a,d}$,Refl$_{c}$,BS$_{b,c}$,Holo$_{a(2)}$;C3;1.0;0.0757;$(3,3,2)$
Refl$_{c}$,Holo$_{d(-1)}$,BS$_{c,d}$,BS$_{b,d}$,DP$_{c,1}$;C3;1.0;0.0756;$(3,3,2)$
DP$_{c,1}$,Refl$_{b}$,Holo$_{b(-1)}$,BS$_{a,b}$,BS$_{a,d}$;C3;1.0;0.0707;$(3,3,2)$
DP$_{a,1}$,BS$_{a,d}$,Refl$_{c}$,BS$_{b,c}$,Refl$_{a}$;C3;0.0;0.0628;$(3,3,2)$
Refl$_{c}$,Holo$_{d(-1)}$,BS$_{c,d}$,BS$_{b,d}$,DP$_{b,1}$;C3;1.0;0.0623;$(3,3,2)$
Holo$_{d(1)}$,Refl$_{c}$,BS$_{c,d}$,BS$_{b,d}$,DP$_{d,1}$;C3;1.0;0.0608;$(3,3,2)$
Refl$_{d}$,Holo$_{d(-1)}$,BS$_{c,d}$,BS$_{b,d}$,DP$_{d,1}$;C3;1.0;0.049;$(3,3,2)$
Refl$_{c}$,Holo$_{d(-1)}$,BS$_{c,d}$,BS$_{b,d}$;C4;1.0;0.0991;$(3,3,2)$
BS$_{b,d}$,BS$_{a,b}$,DP$_{b,1}$,Refl$_{b}$,BS$_{a,b}$;C4;1.0;0.0989;$(3,2,2)$
Refl$_{a}$,Holo$_{b(1)}$,BS$_{a,b}$,BS$_{b,d}$;C4;1.0;0.0796;$(3,3,2)$
Holo$_{d(1)}$,Refl$_{c}$,BS$_{c,d}$,BS$_{b,d}$;C4;1.0;0.0789;$(3,3,2)$
Refl$_{d}$,Holo$_{d(-1)}$,BS$_{c,d}$,BS$_{b,d}$;C4;1.0;0.0778;$(3,3,2)$
Holo$_{c(1)}$,Refl$_{c}$,BS$_{c,d}$,BS$_{b,d}$;C4;1.0;0.0565;$(3,3,2)$
Refl$_{a}$,Holo$_{b(1)}$,BS$_{a,b}$,BS$_{b,d}$,Refl$_{d}$;-1;0.0;0.0822;$(3,3,2)$
\end{filecontents*}
\pgfplotstabletypeset[font=\small,
    col sep=semicolon,
    string type,
    every row no 8/.style={before row=\hline},
    every row no 18/.style={before row=\hline},
    every row no 24/.style={before row=\hline},
    every row no 37/.style={before row=\hline},
    every row no 43/.style={before row=\hline},
    columns/pattern/.style={column name=Pattern, column type={|c|}},
    columns/interestingness/.style={column name=Interestingness, column type={|c|}},
    columns/support/.style={column name=Support, column type={|c|}},
    columns/cohesion/.style={column name=Cohesion, column type={|c|}},
    columns/info/.style={column name=SRV, column type={|c|}},
    every head row/.style={before row=\hline,after row=\hline},
    every last row/.style={after row=\hline},
    ]{Tables/utility_qo_31_wo.csv}
\caption{List of gadgets mined from 10 datasets and clustered by utility. The clusters are separated by horizontal lines and ordered from the highest interestingness to the lowest. Each gadget is assigned a probability (column labeled ``Prob") for belonging to the cluster. If this probability is zero, the sample is considered noise for the corresponding cluster. Cluster -1 contains all gadgets that are not assigned to any of the other clusters. The clusters 1-4 predominately contain interferometers to generate the $(3,3,2)$-state, while the gadgets in cluster 0 generate higher SRV states by using the combination of a hologram and a beam splitter. Within each cluster, the gadgets are ordered by interestingness. Additionally, for each gadget, the support and cohesion are displayed. }
\label{tab:utility_qo}
\end{table}

\newpage

\subsection{Quantum information environment}\label{appendix:qi_env}
Here, we describe in detail the quantum information environment of Sec.~\ref{sec:results_qi}. In this environment, the agent needs to complete an initial quantum circuit without losing the quantum information that is encoded in the first qubit. The 4-qubit circuit that is given to the agent consists of three operations randomly chosen by the environment, placed at fixed positions (marked with an E in Fig.~\ref{fig:envs} right). The agent then completes the circuit by sequentially placing six additional operations in between the initial ones (see Fig.~\ref{fig:envs} right). Thus, the final circuit has 9 operations: 3 placed by the environment and 6 by the agent. The list of possible operations is given in Table~\ref{table:operations_qinf}. Each operation is encoded in 4 features: feature 1 distinguishes between a gate or a measurement (G/M), feature 2 declares a 1-qubit or a 2-qubit operation (1/2), feature 3 labels the register on which the action is applied, and feature 4 labels the specific operation from Table~\ref{table:operations_qinf}. A circuit element is thus described by a 4-feature vector. For example, a Hadamard gate on the third register is denoted as $(G, 1, 3, H)$, or, to ease the notation, $G1_{3H}$. 

The agent's observation is the current state of the circuit, given by the sequence of operations that have already been placed. Each operation is described by four features as explained above. The DDQN receives as input a further encoded vector, where each operation of the circuit is encoded as the concatenation of the one-hot encoding of each feature. 

At each time step, the agent observes the current state of the circuit and chooses an action, i.e., which circuit element to place next. For a 4-qubit circuit, the total number of different possible actions is 58. Following each action, the environment evaluates the circuit and issues a reward. The reward function distinguishes whether the quantum information, which is initially located on the first register, has been lost or not. In order to check whether circuit $C$ keeps the quantum information alive, we verify that the circuit maps orthogonal input states to orthogonal output states. That is, we consider the starting states $|0\rangle|\varphi_i\rangle$, $|1\rangle|\varphi_i\rangle$, where the ancillas can either be initialized in $|\varphi_1\rangle = |000\rangle$ or $|\varphi_2\rangle = |+++\rangle$. If the resulting states $|\psi_a\rangle = C|0\rangle|\varphi_i\rangle$ and $|\psi_b\rangle = C|1\rangle|\varphi_i\rangle$ are nonzero and orthogonal in at least one of the two cases $|\varphi_{i=1,2}\rangle$, the reward is $R=1$. Otherwise, the reward is negative $R=-1$. A reward ($R\in \{1,-1\}$) is issued every time the agent places a new operation on the circuit. Once the agent has completed the full circuit (it has placed 6 operations, completing a 9-operation circuit), it receives a reward that is rescaled by a factor of 5 ($R\in \{5,-5\}$).

\begin{table}[htp]
\centering
\begin{tabular}{|c|c|c|}
\hline
&\textbf{1-qubit} & \textbf{2-qubit}  \\ \hline
\textbf{Gates}      & X, Z, H, S &  CNOT, CZ\\ \hline
\textbf{Measurements} &  $P_X$, $P_Y$, $P_Z$ & $P_{\textrm{Bell}}$,  $ S\otimes \mathbb{I} \, P_{\textrm{Bell}} \, S^\dagger \otimes \mathbb{I}$,  $\mathbb{I} \otimes S \, P_{\textrm{Bell}} \, \mathbb{I} \otimes S^\dagger $\\ \hline

\end{tabular}

\caption{List of available operations for the quantum information environment. $X, Z$ denote the respective Pauli operators, $H$ the Hadamard gate and $S = \sqrt{Z}$ the phase gate. The 2-qubit gates CNOT, CZ are the controlled $X$ and $Z$ gates, respectively. Projectors onto the $+1$ eigenstate of Pauli operator $\sigma=X,Y,Z$ are denoted by $P_\sigma$. Projections onto the Bell state $\frac{1}{\sqrt{2}}(\ket{00}+\ket{11})$ are denoted $P_{\textrm{Bell}}$.} \label{table:operations_qinf}
\end{table}

\begin{figure}[htbp]
\begin{center}
\includegraphics[width=15cm]{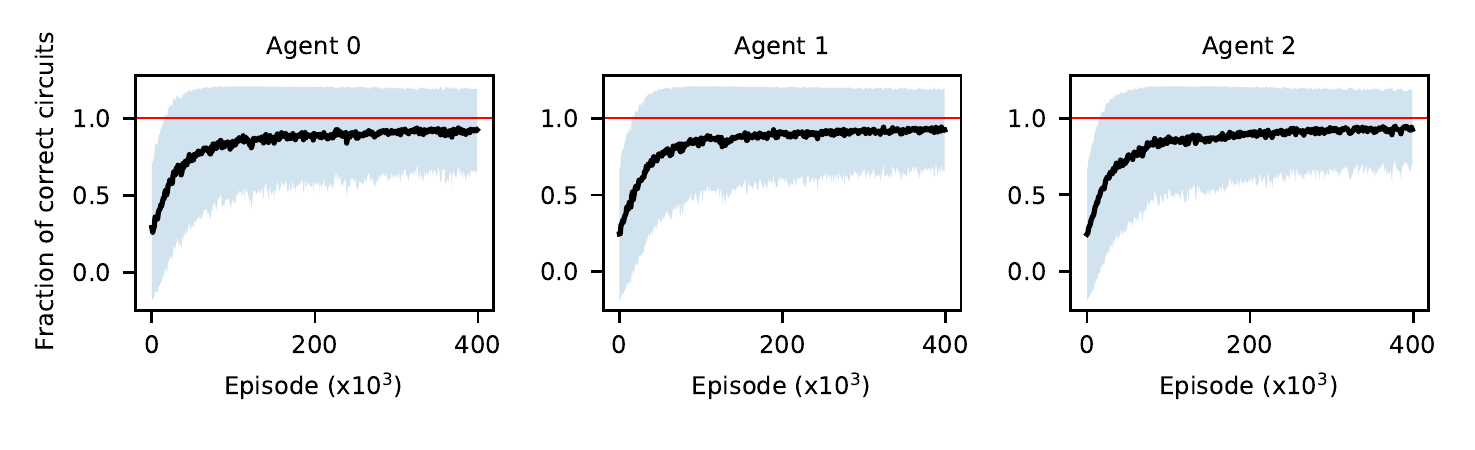}
\caption{\textbf{Learning performance.} Performance (measured as the fraction of circuits that the agent completed successfully) of the three agents trained in the quantum information environment. The agent completes one circuit per episode. Average and one standard deviation over 1000 episodes.}
\label{fig:learning_performance_qi}
\end{center}
\end{figure}

The learning curves of the three independently trained DDQN agents are shown in Fig.~\ref{fig:learning_performance_qi}. The performance is measured as the fraction of circuits that the agent completed successfully. The agent completes one circuit per training episode. Once the agent finishes $4\cdot10^5$ training episodes, the learning process stops and the agent starts collecting positively rewarded circuits until it completes a dataset $D$ with $|D| = 8\cdot10^4$ unique, correct circuits of length 9. This dataset $D$ is then input to the sequence mining (see Sec.~\ref{sec:methods:data_mining}) to mine for gadgets. The gadgets obtained in this environment are listed and analyzed in the following section.

\subsection{Quantum information gadgets}\label{appendix:results_qi}

\textbf{Gadget mining.}
The dataset $D$ of unique, positively rewarded quantum circuits that the agent collected after the training is input to the sequence mining algorithm (see Sec.~\ref{sec:methods:data_mining}) to generate a set of gadgets. From the 9 operations that each circuit contains, we choose to mine only the sequence consisting of the 6 operations, i.e., actions, placed by the agent. In this way, we gain insights about the policy that the agent has learned to complete the circuits effectively.
In addition, as a preprocessing step to the sequence mining, the last feature of each action (i.e., information on the specific gate or measurement basis) in the sequence is removed. The threshold values $F_\textrm{min}$, $C_\textrm{min}$ and $I_\textrm{min}$ are automatically adjusted such that the sequence mining outputs between 1 and 10 gadgets. The resulting values for the thresholds for each agent are given in Table~\ref{table:final_thresholds_qi}.
When selecting gadgets for this environment, we apply two additional filters (see App.~\ref{appendix:dm} for details): (1) Gadgets are filtered by length to allow only gadgets of length greater than 2. (2) Gadgets are filtered by coverage with a maximum coverage value of 2.

\begin{table}[htbp]
\centering
\begin{tabular}{|c|l|l|l|}
\hline
\textbf{Agent id} & \multicolumn{1}{c|}{0}    & \multicolumn{1}{c|}{1}    & \multicolumn{1}{c|}{2}  \\ \hline
\textbf{$F_\textrm{min}$}      & \multicolumn{1}{c|}{0.16} & \multicolumn{1}{c|}{0.20} & \multicolumn{1}{c|}{0.23} \\ \hline
\textbf{$C_\textrm{min}$}      & 0.53                      & 0.61                      & 0.67  \\ \hline

\end{tabular}
\caption{Threshold values for the support $F$ and cohesion $C$ for each agent in the gadget mining. We fix a range between 1 and 10 gadgets that the algorithm should output, and it automatically adjusts the threshold values accordingly. For the minimum interestingness, we set $I_\textrm{min}=F_\textrm{min}\cdot C_\textrm{min}$.}
\label{table:final_thresholds_qi}
\end{table}

\textbf{Gadget clustering.} In order to help us interpret the list of gadgets --collected from three independent DDQN agents--, we cluster them by (i) the type of operations that it contains (see Sec.~\ref{subsec:utility}) and (ii) the context in which they were used (see Sec.~\ref{subsec:context}). In both cases, the clustering is done by the algorithm HDBSCAN (see Sec.~\ref{appendix:HDBSCAN}), and the difference lies in the distance measure that is used. \\

\textit{Utility-based clustering. }As for many reinforcement learning environments, we understand the actions of the agent well enough to feature encode them. In this case, we can categorize actions into four broad types, namely, 1- and 2-qubit gates (denoted $G1$, $G2$) as well as 1- and 2-qubit measurements (denoted $M1$, $M2$). We make use of this coarse-grained information to define a distance measure between gadgets. Each gadget is encoded as a vector of the form $\phi(g) = (n_{G1}, n_{G2}, n_{M1}, n_{M2})$, where each component is the total number of operations of a given type that the gadget contains. We can also define a utility vector $\alpha = (\alpha_{G1}, \alpha_{G2}, \alpha_{M1}, \alpha_{M2})$ that assigns a weight to each operation type. In this case, we assign the same utility to each operation type, i.e. $\alpha = (1,1,1,1)$. The distance between two gadgets $M(g_1, g_2)$ is then defined as the euclidean distance between $\phi(g_1)$ and $\phi(g_2)$. Clustering gadgets with this distance, we obtain the clusters in Table~\ref{tab:utility_qi}. We observe that gadgets with the same types of operations are clustered together. For example, all gadgets that contain at least two 1-qubit gates belong to cluster C1.\\


\begin{table}
\centering
\begin{filecontents*}{Tables/utility_clusters_new.csv}
Gadget;Cluster;Prob;Interestingness;Support;Cohesion
$G2_{23}$,$G1_{2}$;C2;1;0.3096;0.3269;0.9473
$G2_{24}$,$G1_{2}$,$G1_{2}$;C2;1;0.2254;0.3143;0.7172
$G1_{4}$,$G2_{24}$;C2;1;0.2186;0.219;0.9983
$G2_{34}$,$G1_{4}$;C2;1;0.2297;0.2306;0.9958
$G2_{34}$,$G1_{1}$,$G1_{1}$;C2;1;0.22;0.2912;0.7556
$G1_{2}$,$G2_{34}$;C2;1;0.2188;0.2831;0.7729
$G1_{4}$,$G1_{4}$;C1;1;0.4536;0.4771;0.9508
$G1_{1}$,$G1_{4}$;C1;1;0.3557;0.4914;0.7239
$G1_{3}$,$G1_{4}$;C1;1;0.3358;0.3719;0.903
$G1_{1}$,$G1_{3}$;C1;1;0.2799;0.3085;0.9073
$G1_{4}$,$G1_{4}$,$G1_{4}$;C1;0;0.2525;0.2653;0.9515
$G1_{2}$,$G1_{2}$;C1;1;0.4707;0.476;0.9889
$G1_{2}$,$G1_{2}$,$G1_{2}$;C1;0;0.3054;0.3101;0.9847
$G1_{1}$,$G1_{1}$;C1;1;0.4445;0.4885;0.9098
$G1_{4}$,$G1_{1}$;C1;1;0.2417;0.269;0.8986
$G1_{2}$,$G1_{3}$;C1;1;0.2285;0.2568;0.8897
$G1_{3}$,$G1_{1}$,$G1_{1}$;C1;0;0.2011;0.2582;0.7791
$M2_{34}$,$M2_{34}$;C0;1;0.2192;0.2276;0.9631
$G1_{4}$,$M1_{2}$;C0;0.71;0.2085;0.2794;0.7461
$G2_{12}$,$M2_{34}$;C0;0.71;0.2012;0.2302;0.874
$G2_{12}$,$M2_{13}$;C0;0.71;0.1454;0.233;0.6239
$G2_{12}$,$M2_{14}$;C0;0.71;0.1281;0.1623;0.7893
$G2_{24}$,$G2_{23}$;C0;0.71;0.4394;0.4462;0.9849
$G2_{24}$,$G2_{24}$;C0;0.71;0.27;0.2851;0.9471
$M2_{14}$,$M2_{14}$;C0;1;0.2455;0.2557;0.9601
$G2_{23}$,$G2_{23}$;C0;0.71;0.227;0.2288;0.9924
$M2_{34}$,$M2_{13}$;C0;1;0.3138;0.3233;0.9708
$M2_{34}$,$M2_{13}$,$M2_{12}$;C0;1;0.2476;0.2494;0.9929
$M2_{13}$,$M2_{12}$;C0;1;0.2472;0.2494;0.9913
\end{filecontents*}
\pgfplotstabletypeset[font=\small,
    col sep=semicolon,
    string type,
    every row no 6/.style={before row=\hline},
    every row no 17/.style={before row=\hline},
    columns/Gadget/.style={column name=Gadget, column type={c}},
    columns/Cluster/.style={column name=Cluster, column type={c}},
    columns/Prob/.style={column name=Prob., column type={c}},
    columns/Interestingness/.style={column name=Interestingness, column type={c}},
    columns/Support/.style={column name=Support, column type={c}},
    columns/Cohesion/.style={column name=Cohesion, column type={c}},
    every head row/.style={before row=\hline,after row=\hline},
    every last row/.style={after row=\hline},
    ]{Tables/utility_clusters_new.csv}

\caption{List of gadgets obtained by the three DDQN agents and classified through utility-based clustering. The column ``Prob" indicates the probability that the gadget belongs to the given cluster. A zero probability means that this data point is considered noise. For each gadget, we also provide its interestingness, support, and cohesion. The last gadgets are unclassified. }
\label{tab:utility_qi}
\end{table}

\textit{Context-based clustering. }We define another distance between two gadgets as the distance between their corresponding initialization sets (see Sec.~\ref{subsec:context}). First, the initialization set $\mathcal{I}_g$ of each gadget $g$ is obtained from the set of interaction sequences $D_g$ that contain $g$. $\mathcal{I}_g$ is the set of observations prior to the placement of gadget $g$. We collect initialization sets of size $|\mathcal{I}_{g_1}|=|\mathcal{I}_{g_2}|=2000$. Then, the distance $m(o,q)$ for each pair of observations $o \in \mathcal{I}_{g_1}$ and $q \in \mathcal{I}_{g_2}$ is computed as
\begin{equation}
    m(o,q) = \frac{1}{L} \sum_{k=1}^L \sum_{i=1..3} w_i \, d(a_i^{(k)}, b_i^{(k)}),
\end{equation}
where observations $o$ and $q$ are defined as sequences of actions $o = (a^{(1)}, ..., a^{(L)})$, $q = (b^{(1)}, ..., b^{(L)})$. Note that only sequences of the same length $|o|=|q|=L$ are compared. 
Each action $a^{(k)}$ in the sequence is an operation described as a feature vector $a^{(k)} = (a^{(k)}_1, a^{(k)}_2, a^{(k)}_3)$ (the last feature was removed in the gadget mining). We then choose $d(a_i^{(k)}, b_i^{(k)})$ as the normalized Hamming distance between the feature $i$ of operations $a^{(k)}$ and $b^{(k)}$. Analogous to the utilities defined in Sec.~\ref{subsec:context}, the user can introduce weights $w_i$ for each feature $i$. 
In this case, we choose equally distributed weights. The distance $M(\mathcal{I}_{g_1},\mathcal{I}_{g_2})$ that is used to cluster gadgets $g_1$ and $g_2$ is then computed according to Eq.~\ref{eq:distance_initsets}. 

\begin{figure}[htbp]
\begin{center}
\includegraphics[width=8cm]{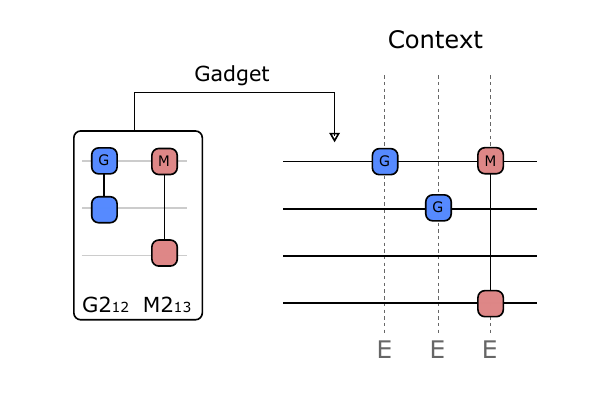}
\caption{\textbf{Teleportation gadget.} Example of a circuit in the context of gadget $G2_{12}, M2_{13}$. Before placing the gadget, the circuit yields a negative reward ($R=-1$) due to the measurement on the first register. After the agent places the gadget, the circuit is positively rewarded because the quantum information is teleported.}
\label{fig:change_reward_after_gadget_qi}
\end{center}
\end{figure}

Table~\ref{tab:context_qi} shows the classification by context. In this case, we can gain insights into the policy and environment by analyzing not only the clusters and gadgets themselves but also the initialization sets of the corresponding gadgets. In Table~\ref{tab:context_qi}, we provide information on (i) the average length of the circuits inside $\mathcal{I}_{g}$ for each $g$, which indicates at which point of the circuit the agent decides to place the gadget on average; and (ii) the average reward of circuits in $\mathcal{I}_g$, which informs us about whether the agent places the gadget to correct the circuit (negative InitSet Reward) or not. 
By visually inspecting some circuits in the initialization sets of gadgets in cluster C2 with negative InitSet Reward, we see that, as expected, they contain measurements on the first register (see e.g., Fig.~\ref{fig:change_reward_after_gadget_qi}). We confirmed that the agent learned to use gadgets in C2 to correct the circuit by analyzing the reward of the circuit before and after placing one of these gadgets, which turned from negative to positive (see Fig.~\ref{fig:change_reward_after_gadget_qi} for an example of such a case). In addition, we observe that the circuits in the initialisation set $\mathcal{I}_g$ of gadgets $g$ that belong to cluster C0 and of the unclassified gadgets are longer on average. Again by inspecting the circuits in the corresponding $\mathcal{I}_g$, we see that almost all start with a 1-qubit gate (either $G1_1$ or $G1_2$) for $g$ in cluster C0. An example can be seen in Fig.~\ref{fig:results_qi_context}: the circuit displayed in the label of C0 is an element of the initialization set of gadget $(G1_4, G1_4)$, where the first operation is $G1_1$. $80.7\%$ of the circuits in its context start with $G1_1$. This percentage is higher for other gadgets in C0, for example $(G1_4, M1_2)$, for which $98.9\%$ of the circuits in its initialization set start with $G1_1$. Table~\ref{tab:initial_operation_contexts_qi} shows the percentages of circuits that start with a given operation in the initialization set of gadgets belonging to C0, C1, and the unclassified gadgets. This explains why gadgets in C0 are clustered separately.

\begin{table}
\centering
\begin{filecontents*}{Tables/gadget_clusters_all_info_Hamming_distance_new.csv}
Gadget;Cluster;Prob;InitSet Length;InitSet Reward;Interestingness;Agent 
$G2_{12}$,$M2_{34}$;C2;1;3.00;-0.64;0.2012;0
$G2_{12}$,$M2_{13}$;C2;1;3.00;-0.72;0.1454;0
$G2_{12}$,$M2_{14}$;C2;1;3.00;-0.76;0.1281;0
$M2_{34}$,$M2_{13}$;C2;1;3.00;-0.76;0.3138;2
$M2_{34}$,$M2_{13}$,$M2_{12}$;C2;1;3.00;-0.75;0.2476;2
$G1_{1}$,$G1_{4}$;C1;1;3.02;0.99;0.3557;0
$G1_{1}$,$G1_{3}$;C1;1;3.00;0.99;0.2799;0
$G2_{24}$,$G2_{23}$;C1;0.996;3.51;0.99;0.4394;1
$G2_{24}$,$G2_{24}$;C1;0.996;3.21;0.98;0.27;1
$G2_{24}$,$G1_{2}$,$G1_{2}$;C1;0.877;3.62;0.99;0.2254;1
$G1_{4}$,$G2_{24}$;C1;1;3.08;0.99;0.2186;1
$G1_{2}$,$G1_{3}$;C1;1;3.12;0.99;0.2285;2
$G1_{2}$,$G2_{34}$;C1;1;3.08;0.98;0.2188;2
$G1_{4}$,$G1_{4}$;C0;1;4.80;0.84;0.4536;0
$G1_{3}$,$G1_{4}$;C0;1;4.42;0.82;0.3358;0
$G1_{4}$,$G1_{4}$,$G1_{4}$;C0;1;4.36;0.88;0.2525;0
$G1_{4}$,$M1_{2}$;C0;1;4.71;0.97;0.2085;0
$G1_{4}$,$G1_{1}$;C0;0.876;5.32;0.96;0.2417;2
$G2_{34}$,$G1_{4}$;C0;0.837;5.01;0.95;0.2297;2
$G2_{34}$,$G1_{1}$,$G1_{1}$;C0;0.846;5.00;0.95;0.22;2
$G1_{3}$,$G1_{1}$,$G1_{1}$;C0;1;4.77;0.97;0.2011;2
$M2_{34}$,$M2_{34}$;-1;0;5.33;0.22;0.2192;0
$G1_{2}$,$G1_{2}$;-1;0;6.00;0.72;0.4707;1
$G2_{23}$,$G1_{2}$;-1;0;4.98;0.98;0.3096;1
$G1_{2}$,$G1_{2}$,$G1_{2}$;-1;0;5.70;0.67;0.3054;1
$M2_{14}$,$M2_{14}$;-1;0;4.16;0.00;0.2455;1
$G2_{23}$,$G2_{23}$;-1;0;4.71;0.99;0.227;1
$G1_{1}$,$G1_{1}$;-1;0;6.02;0.97;0.4445;2
$M2_{13}$,$M2_{12}$;-1;0;4.00;-0.08;0.2472;2
\end{filecontents*}
\pgfplotstabletypeset[font=\small,
    col sep=semicolon,
    string type,
    every row no 5/.style={before row=\hline},
    every row no 13/.style={before row=\hline},
    every row no 21/.style={before row=\hline},
    columns/Pattern/.style={column name=Gadget, column type={c}},
    columns/Cluster/.style={column name=Cluster, column type={c}},
    columns/Prob/.style={column name=Prob., column type={c}},
    columns/Initset Length/.style={column name=Initset Length, column type={c}},
    columns/Initset Reward/.style={column name=Initset Reward, column type={c}},
    columns/Interestingness/.style={column name=Interestingness, column type={c}},
    columns/Agent/.style={column name=Agent, column type={c}},
    every head row/.style={before row=\hline,after row=\hline},
    every last row/.style={after row=\hline},
    ]{Tables/gadget_clusters_all_info_Hamming_distance_new.csv}
    
\caption{List of gadgets obtained by the three DDQN agents classified by context. The column ``Prob" indicates the probability that the gadget belongs to the given cluster. The columns ``InitSet Length" and ``InitSet Reward" indicate the average length and average reward, respectively, of the circuits in the initialization set of the gadget. The column labeled "Agent" shows from which agent's dataset the gadget was retrieved. The last gadgets are unclassified. }
\label{tab:context_qi}
\end{table}

\begin{table}
\centering
\begin{filecontents*}{Tables/gadget_clusters_first_operation_new.csv}
Gadget;Cluster;$G1_{1}$;$G1_{2}$;$G1_{4}$;$G2_{12}$;$G2_{14}$;$G2_{23}$;$G2_{24}$;$M2_{34}$
$G2_{24}$,$G2_{23}$;C1;0.001;0.002;0.392;0.001;0;0.074;0;0
$G2_{24}$,$G2_{24}$;C1;0;0.001;0.156;0;0;0.018;0;0
$G2_{24}$,$G1_{2}$,$G1_{2}$;C1;0.002;0;0.569;0.001;0.002;0.012;0;0.001
$G1_{4}$,$G1_{4}$;C0;0.807;0.001;0;0.121;0;0.066;0;0
$G1_{3}$,$G1_{4}$;C0;0.726;0;0;0.212;0.001;0.059;0;0.0005
$G1_{4}$,$G1_{4}$,$G1_{4}$;C0;0.882;0.001;0;0.046;0;0.067;0;0.0015
$G1_{4}$,$M1_{2}$;C0;0.989;0.001;0;0.003;0;0.004;0;0
$G1_{4}$,$G1_{1}$;C0;0.214;0.688;0;0.001;0.001;0.006;0.002;0.0015
$G2_{34}$,$G1_{4}$;C0;0.213;0.721;0.009;0;0;0;0.001;0
$G2_{34}$,$G1_{1}$,$G1_{1}$;C0;0.208;0.732;0.007;0.001;0.001;0;0.001;0
$G1_{3}$,$G1_{1}$,$G1_{1}$;C0;0.235;0.676;0.001;0.001;0.001;0.002;0;0.0005
$M2_{34}$,$M2_{34}$;-1;0.414;0.001;0.001;0.584;0.001;0.001;0;0
$G1_{2}$,$G1_{2}$;-1;0.001;0;0.381;0.017;0.315;0.011;0.25;0.0015
$G2_{23}$,$G1_{2}$;-1;0;0.001;0.49;0.001;0.002;0;0.447;0.0005
$G1_{2}$,$G1_{2}$,$G1_{2}$;-1;0.002;0;0.357;0.008;0.382;0.006;0.234;0.0005
$M2_{14}$,$M2_{14}$;-1;0;0;0;0.948;0.053;0;0;0
$G2_{23}$,$G2_{23}$;-1;0;0.001;0.229;0.002;0;0;0.581;0
$G1_{1}$,$G1_{1}$;-1;0;0.54;0.007;0;0.001;0.001;0;0.165
$M2_{13}$,$M2_{12}$;-1;0;0;0;0;0;0;0;1
\end{filecontents*}
\pgfplotstabletypeset[font=\small,
    col sep=semicolon,
    string type,
    every row no 3/.style={before row=\hline},
    every row no 11/.style={before row=\hline},
    columns/Pattern/.style={column name=Pattern, column type={c}},
    columns/Cluster/.style={column name=Cluster, column type={c}},
    columns/$G1_1$/.style={column name=$G1_1$, column type={c}},
    columns/$G1_2$/.style={column name=$G1_2$, column type={c}},
    columns/$G1_4$/.style={column name=$G1_4$, column type={c}},
    columns/$G2_{12}$/.style={column name=$G2_{12}$, column type={c}},
    columns/$G2_{14}$/.style={column name=$G2_{14}$, column type={c}},
    columns/$G2_{23}$/.style={column name=$G2_{23}$, column type={c}},
    columns/$G2_{24}$/.style={column name=$G2_{24}$, column type={c}},
    columns/$M2_{34}$/.style={column name=$M2_{34}$, column type={c}},
    every head row/.style={before row=\hline,after row=\hline},
    every last row/.style={after row=\hline},
    ]{Tables/gadget_clusters_first_operation_new.csv}
    
\caption{Fraction of circuits that start with the same operation (indicated in the column label) in the initialization set of gadgets that belong to clusters C0 and C1. Table displays gadgets with initialization sets that have an average length larger than 3. Only the most frequent starting operations are listed.}
\label{tab:initial_operation_contexts_qi}
\end{table}

\newpage

\section{Implementation details}
For the quantum optics environment, we use an intrinsically motivated MCTS, with horizon $T=12$, $C=0.01$ that balances exploration and exploitation, and a boredom threshold of $B=1000$ time steps after which exploitative agents are reset.

 For the quantum information environment, we showcase that a standard implementation of a deep RL agent can be integrated and analysed with the present architecture. In particular, we use the DDQN implementation from the  {\href{https://github.com/mostaszewski314/rl_for_vqe}{github repository}} accompanying the paper \cite{Ostaszewski2021}. The neural network for the implementation has three layers with 150 neurons each. We choose the ADAM optimizer with a learning rate of $0.00001$. The training batches of the main network are of size 100 and the capacity of the replay memory is of 2000 interactions, updated in a first-in first-out manner. The target network is updated every 1000 calls to the replay memory, which occurs every time step once the first batch is filled. The agent's policy is $\epsilon$-greedy, with $\epsilon$ decaying exponentially with factor $0.999991$, until a minimum value $\epsilon = 0.01$. 

Details on the sequence mining parameters and filters chosen for the quantum optics and quantum information environments are given in appendices~\ref{appendix:results_qo} and \ref{appendix:results_qi}, respectively.

For clustering, we applied HDBSCAN with a minimum number of samples of 2 and a minimum cluster size of 5, in all cases.

For further implementation details, we refer to {\href{https://github.com/LeaMarion/automated_gadget_discovery.git}{our open-source repository on GitHub}}.




\end{document}